\documentclass[journal]{IEEEtran}

\usepackage{algorithm, algorithmic}

\usepackage{array}
\usepackage{subfigure}
\usepackage{textcomp}
\usepackage{stfloats}
\usepackage{url}
\usepackage{verbatim}
\usepackage{hyperref}					
\hypersetup{colorlinks,
	linkcolor=blue,%
	anchorcolor=blue,
    citecolor=blue}

\usepackage{bm}

\usepackage{cite}
\usepackage{amsmath,amssymb,amsfonts}

\usepackage{amsthm}
\newtheorem{proposition}{Proposition}

\newtheorem{corollary}{Corollary}

\newtheorem{assumption}{Assumption}

\usepackage{algorithmic}
\usepackage{graphicx}

\usepackage{textcomp}
\usepackage{xcolor}

\usepackage{array} 

\usepackage{flushend}

\begin{document}
	
	\title{\huge Reshaping the ISAC Tradeoff Under OFDM Signaling:\\A Probabilistic Constellation Shaping Approach}

	\author{Zhen Du,~\IEEEmembership{Member,~IEEE,} Fan Liu,~\IEEEmembership{Senior Member,~IEEE,} Yifeng Xiong,~\IEEEmembership{Member,~IEEE,} \\Tony Xiao Han,~\IEEEmembership{Senior Member,~IEEE,} Yonina C. Eldar,~\IEEEmembership{Fellow,~IEEE}, and~Shi Jin,~\IEEEmembership{Fellow,~IEEE} 
        \thanks{(\textit{Corresponding author: Fan Liu})}
		\thanks{An earlier version was partly presented at the IEEE Global Communications Conference (GLOBECOM), Kuala Lumpur, Malaysia, Dec 2023 \cite{PCSglobecom}. }
		\thanks{Z. Du is with the School of Electronic and Information Engineering, Nanjing University of Information Science and Technology, Nanjing, China. 
		}
		\thanks{F. Liu is with School of System Design and Intelligent Manufacturing, Southern University of Science and Technology, Shenzhen, China. 
		}
		\thanks{Y. Xiong is with the School of Information and Electronic Engineering,	Beijing University of Posts and Telecommunications, Beijing, China. 
        }
		\thanks{T. X.-Han is with Huawei Technologies Co., Ltd, Shenzhen, China. 
		}
		\thanks{Y. C. Eldar is with the Faculty of Mathematics and Computer Science, Weizmann Institute of Science, Rehovot, Israel. 
		}
            \thanks{S. Jin is with National Mobile Communications Research Laboratory, Southeast University, Nanjing, 210096, China.
            }
	}
	
	

	\maketitle

	\begin{abstract}
		Integrated sensing and communications is regarded as a key enabling technology in the sixth generation networks, where a unified waveform, such as orthogonal frequency division multiplexing (OFDM) signal, is adopted to facilitate both sensing and communications (S\&C). However, the random communication data embedded in the OFDM signal results in severe variability in the sidelobes of its ambiguity function (AF), which leads to missed detection of weak targets and false detection of ghost targets, thereby impairing the sensing performance. Therefore, balancing between preserving communication capability (i.e., the randomness) while improving sensing performance remains a challenging task. To cope with this issue, we characterize the random AF of OFDM communication signals, and demonstrate that the AF variance is determined by the fourth-moment of the constellation amplitudes. 
		Subsequently, we propose an optimal probabilistic constellation shaping (PCS) approach by maximizing the achievable information rate (AIR) under the fourth-moment, power and probability constraints, where the optimal input distribution may be numerically specified through a modified Blahut-Arimoto algorithm. To reduce the computational overheads, we further propose a heuristic PCS approach by actively controlling the value of the fourth-moment, without involving the communication metric in the optimization model, despite that the AIR is passively scaled with the variation of the input distribution. 
		Numerical results show that both approaches strike a scalable performance tradeoff between S\&C, where the superiority of the PCS-enabled constellations over conventional uniform constellations is also verified. Notably, the heuristic approach achieves very close performance to the optimal counterpart, at a much lower computational complexity. 
	\end{abstract}
	

	\section{Introduction}
	\IEEEPARstart{W}{ireless} sensing is envisioned as a native capability of sixth generation (6G) networks, which facilitates many emerging applications requiring reliable and high-precision location-aware functionality \cite{liu2023integrated,cui2021integrating,liu2022integrated}. This triggers the recent development of integrated sensing and communications (ISAC) technology that has received official recognition from the international telecommunication union (ITU) \cite{ITU-R}, as one of the vertices supporting the 6G hexagon of usage scenarios. Indeed, ISAC enables a synergistic design of previously isolated sensing and communication (S\&C) functionalities, which not only improves the utilization efficiency of both hardware and wireless resources, but may also lead to mutual performance gains between S\&C \cite{du2023towards,dong2023communication}.
	
	To fully realize the promise of ISAC technologies, a unified signal capable of simultaneously accomplishing both S\&C tasks is indispensable. In general, ISAC signaling strategies may be classified into three design philosophies, namely, sensing-centric \cite{saddik2007ultra,hassanien2015dual,huang2020majorcom} and communication-centric \cite{sturm2011waveform,kumari2017ieee} designs that are built upon legacy S\&C waveforms, respectively, and joint designs \cite{liu2018mu} aiming to conceive ISAC waveforms from the ground-up. While the joint design strategy may potentially achieve the Pareto performance boundary, it suffers from high computational complexity and lack of compatibility with existing S\&C infrastructures. To that end, communication-centric designs may be a more viable and low-cost solution to implement ISAC in 6G wireless networks, where standardized orthogonal frequency division multiplexing (OFDM) waveforms can be straightforwardly adopted for target sensing. 
	
	The feasibility of employing OFDM signals for target detection \cite{sen2010adaptive}, estimation \cite{sen2012ofdm}, and tracking \cite{sen2010ofdm} has been verified for radar systems in the last decade.  
	Along this line of research, OFDM radar waveform optimization has been well-investigated for enhancing the sensing performance \cite{bicua2016generalized,shi2017power,bicua2018radar,du2020distributed}.
	However, most existing works treat OFDM radar signals as deterministic, where random communication symbols are replaced by deterministic weights representing power allocation across subcarriers. Consequently, the sensing performance is optimized at the price of elimination of the signaling randomness, i.e., the loss of data transmission capability.
	
	Despite the well-evolved OFDM radar theory, applying the OFDM framework to ISAC systems still faces several challenges. To convey useful information, OFDM-based ISAC signals have to accommodate randomness of the communication data. In particular, each subcarrier is endowed with a communication symbol randomly drawn from certain codebooks, e.g., phase shift keying (PSK) or quadrature amplitude modulation (QAM) constellations. Highly random signals ensure high-throughput transmission, but may jeopardize target sensing. As evidenced in \cite{xiong2023fundamental}, random Gaussian signals, known to be capacity-achieving for point-to-point Gaussian channels, lead to sensing performance loss due to increased Cram\'er-Rao bound (CRB) for parameter estimation, implying a deterministic-random tradeoff between S\&C in terms of their different preference on the input distribution. 
	
	In order to cope with the above issue, exploiting OFDM communication signals for sensing while preserving its randomness has received recent attention \cite{berger2010signal,hakobyan2017novel,zhang2020joint,keskin2021mimo}. The basic rationale is to extract delay and Doppler parameters with the standard matched filtering approach, which may be split into four steps: 1) Compensation for Doppler phase shifts in the time domain; 2) Fast Fourier transform (FFT) across each subcarrier; 3) Compensation of the known data symbols; 4) Inverse FFT across various delays. In this context, delay and Doppler estimation is decoupled, where a target is declared to be present if a peak is detected in the corresponding delay-Doppler resolution cell when comparing with a threshold. A desired matched filter requires adherence to the constant modulus constraint on the transmit waveform, which ensures a flat spectrum and facilitates narrow mainlobes and low sidelobes. PSK-modulated OFDM signals with randomly varying phases and a fixed amplitude typically fulfill this requirement, while QAM fails to meet this criterion due to randomly varying amplitudes. Consequently, adopting QAM-modulated OFDM communication signals directly for target detection leads to compromised sensing performance.

	More relevant to this work, a data division approach has been proposed in \cite{sturm2011waveform}, since data symbols are perfectly known at the sensing receiver in a monostatic ISAC system. By sampling at each OFDM symbol and performing block-wise FFT, 
	the communication symbol is mitigated by element-wise division. However, this operation may change the statistical characteristics of the noise across subcarriers and OFDM symbols, leading to performance degradation in thresholding and peak detection in the subsequent 2D-FFT processing. As a result, PSK modulation is preferable for sensing comparing to QAM, since it keeps the noise distribution unaltered, at the price of reduced communication rate. Indeed, the inherent data randomness results in fluctuated sidelobes of the ambiguity function (AF) for OFDM communication signals, both for matched filtering and data division approaches, leading to missed detection of weak targets and false detection of ghost targets. Thus, the state-of-the-art OFDM signaling schemes are generally unable to balance between S\&C performance, due to the absence of sufficient degrees-of-freedom (DoFs) for controlling the signaling randomness.

	Inspired by the deterministic-random tradeoff in ISAC systems\cite{xiong2023fundamental}, we aim to explore a novel DoF beneficial for bridging between S\&C performance, i.e., the input distribution of constellation symbols, which is different from classic PSK and QAM with uniformly distributed symbols.
	In practice, optimizing the input distribution has been leveraged merely for communication purpose, i.e., minimizing the gap between the achievable information rate (AIR) and channel capacity, which is known as probabilistic constellation shaping (PCS) \cite{bocherer2014probabilistic,bocherer2015bandwidth,cho2019probabilistic,steiner2020coding,schulte2015constant}.
	Nevertheless, the benefits of exploiting PCS in ISAC systems have not yet been clarified and validated.
	
	To be more specific, in this paper our aim is to preserve the communication capability (i.e., the randomness) of OFDM signaling modulated by non-uniformly distributed QAM symbols, while improving its sensing ability relative to a classic QAM counterpart. 
	To this end, we propose a tailored PCS approach to generate random symbols with optimized input probabilities in accordance with S\&C requirements. The core idea here is to build up the exact relationship between the input distribution of random symbols and S\&C metrics, thereby scaling S\&C behaviors with the PCS approach. For clarity, our contributions are summarized as follows:
	\begin{itemize}
		\item We commence by studying the expectation and variance of AF for OFDM communication signals, where i.i.d. input symbols are drawn from a finite complex-valued alphabet known as a constellation, with arbitrary input distribution. Previous works, such as \cite{tigrek2012ofdm}, investigated these properties only for PSK-modulated OFDM signals. We further extend the statistical analysis to OFDM sequences. Our analysis reveals that the variance of AF is closely related to the fourth-order moment of random constellation amplitudes.
		\item We propose to maximize the AIR constrained by the fourth-moment of constellation symbols through leveraging the PCS method. The resulting model is highly nonlinear. To tackle this challenge, we introduce a modified Blahut-Arimoto (MBA) algorithm that is proved to attain the globally optimal distribution. Simulations depict its ability to strike a controllable tradeoff between S\&C performance.
		\item To reduce the computational overheads, we further present a heuristic PCS approach by actively controlling the value of the fourth-order moment, without explicitly involving the AIR constraint. While the model omits the AIR, we show that it achieves a near-optimal S\&C performance tradeoff at a significantly reduced computational complexity. This attributes to the adjustable signaling randomness arisen by optimizing the fourth-moment, thereby balancing between S\&C behaviors.
	\end{itemize}
    
    Numerical results verify that both PCS approaches may attain a flexible tradeoff between S\&C performance in ISAC systems, in contrast to conventional uniformly distributed PSK/QAM constellations that only achieve the sensing/communication-optimal performance, i.e., two corner points on the S\&C tradeoff curve. The tailored algorithms can be implemented offline, indicating their potentials of being efficiently deployed in practical scenarios.
	
    The remainder of this article is organized as follows. Section \ref{sec2} formulates the AF of OFDM signaling and its statistical characteristics. Section \ref{sec3} analyzes the communication metric, i.e., the AIR. Two PCS approaches are proposed in Section \ref{sec5}. Simulations in Section \ref{sec6} demonstrate the advantage of the approaches in striking a flexible tradeoff between S\&C performance. Finally, the article is concluded in Section \ref{sec7}.

	\textit{Notation:} Throughout the paper, $\mathbf{A}$, $\mathbf{a}$, and $a$ denote a matrix, vector, and scalar, respectively. 
	We use $\Re(\cdot)$, $\mathbb{E}(\cdot)$, $\vert\cdot\vert$, $\Vert\cdot\Vert$, $(\cdot)^{T}$, $(\cdot)^{*}$, $(\cdot)^{H}$, $(\cdot)^{-1}$, $\mathbf{1}_{N}$, and $\mathbf{I}_{N}$ to denote the real part of a complex number, expectation, modulus of a complex number, Frobenius norm, transpose, conjugate, Hermitian, inverse, identity vector of size $N \times 1$, and identity matrix of size $N\times N$, respectively.

	\section{Sensing Performance Evaluation}\label{sec2}
	In this section, we commence with the ISAC signal model for a single OFDM symbol. Subsequently, the statistical characteristics of its AF, i.e., the expectation and variance, are derived. Then the more general case of OFDM sequences is analyzed. This lays a foundation for the proposed two PCS approaches in Section \ref{sec5}.

    \subsection{Characterization of AF}	

\begin{figure}[!t]
		\centering
		\includegraphics[width=2.20in]{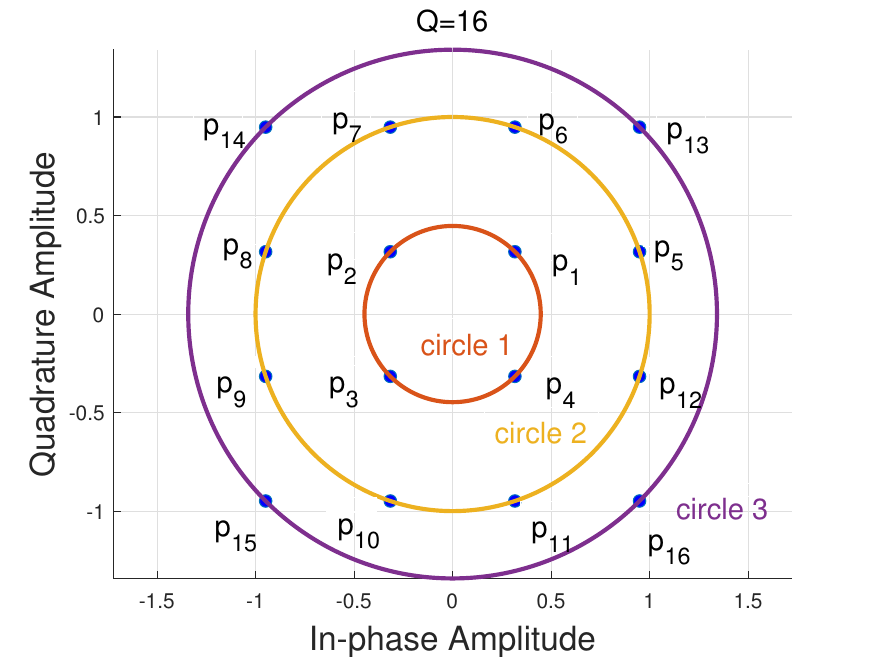}\\
		\caption{QAM constellation: Geometrical and probabilistic symmetry.}
		\label{fig1x}
	\end{figure}

	We consider a monostatic ISAC system employing OFDM signals for S\&C tasks simultaneously. An OFDM symbol consisting of $L$ subcarriers, occupying a bandwidth of $B$ Hz and a symbol duration of $T_p$ seconds, is given by
	\begin{equation}
		\begin{aligned}
			s(t) = \sum\nolimits^{L-1}_{l=0}A_l \exp\left(j\psi_l\right) \exp\left(j2\pi l \Delta ft\right) \text{rect}\left( {t}/{T_p} \right),
		\end{aligned}
	\end{equation}
	where $\exp\left(j2\pi l \Delta f t\right)\triangleq\phi_l(t)$ denotes the $l$th subcarrier, $\Delta f = B/L=1/T_p$ represents the subcarrier interval in the frequency domain,
	$A_l$ and $\psi_l$ denote the amplitude and phase of the $l$th i.i.d. input symbol drawn from a finite complex-valued alphabet, such as the QAM based constellation in Fig. \ref{fig1x}, and 
	$\text{rect}(t)$ represents the rectangle window, equal to $1$ for $0\leq t\leq 1$, and zero otherwise. 

	The randomness of OFDM communication signals lies in the discrete random variables $A_l\exp\{\psi_l\}$ mapped from the bit streams to the constellation symbols with a certain input distribution.	To illustrate this in Fig. \ref{fig1x}, we denote the complex random variable in an arbitrary constellation by $x=A_x \exp\left(\psi_x\right)$, and assume that there are $Q$ discrete constellation points in the given input alphabet $x\in \mathcal{X}=\{x_1,x_2,\cdots,x_Q\}$, where the $q$th point $x_q$ is transmitted with prior probability $p_q$, satisfying $\sum_x p(x) = \sum^{Q}_{q=1} p_q=1$. As a special case, conventional PSK and QAM constellations are uniformly distributed, i.e., $p(x) = \frac{1}{Q}, \ \forall x$. The expectation $\mathbb{E}_Y\{y\}=\sum_x f(x)p(x)$ refers to the summation of $Q$ values weighted by their discrete probabilities in the constellation, where $y=f(x)$ represents a function of the random constellation points $x$. For convenience, we omit the subscript of $\mathbb{E}_Y\{y\}$ in the following. With this definition, the normalized transmit power can be expressed as $\mathbb{E}\{A^2_l\} = 1$.

	Suppose that a target to be detected has delay $\tau_0$ and Doppler shift $\nu_0$. Then the received signal is expressed as
	\begin{equation}
		\begin{aligned}
            y_r(t) = \alpha s(t-\tau_0)\exp\left(j2\pi \nu_0t\right)+n_r(t),
		\end{aligned}
	\end{equation}
    where $\alpha$ and $n_r(t)$ represent the radar cross section (RCS) and input jamming, respectively.
	The matched filtering at the sensing receiver is thus given as
	\begin{equation}
		\begin{aligned}
			\Delta(\tau,\nu) & =  \int^{\infty}_{-\infty} y_r(t)s^*(t-\tau) \exp\left(-j2\pi\nu t\right)dt 
			\\ &  =  \alpha \Lambda\left(\tau-\tau_0,\nu-\nu_0\right) + n'_r(t),
		\end{aligned}
	\end{equation}
	where $\Lambda\left(\tau,\nu\right)$ is followed from the Woodward's definition of AF \cite{woodward2014probability}, which is interpreted as the matched filter response, expressed as a two-dimensional correlation between the transmit signal $s(t)$ and its time-delayed/frequency-shifted counterpart, i.e.,
	\begin{equation}\label{Eq1}
		\begin{aligned}
			\Lambda(\tau,\nu) = & \int^{\infty}_{-\infty} s(t)s^*(t-\tau) \exp\left(-j2\pi\nu t\right)dt.
		\end{aligned}
	\end{equation}

	\begin{proposition}
		The AF of a single OFDM symbol is
		\begin{equation}
			\begin{aligned}
				\Lambda(\tau,\nu) = \Lambda_\text{S}(\tau,\nu) +\Lambda_\text{C}(\tau,\nu),
			\end{aligned}
		\end{equation}
		where
		\begin{equation*}
			\begin{aligned}
				\Lambda_\text{S}(\tau,\nu) & =  T_\text{diff} \text{sinc}\left( -\nu T_\text{diff} \right) \exp\left\{-j2\pi \nu T_\text{avg}\right\} \\ & \times \sum\nolimits^{L-1}_{l=0} A^2_{l}   \exp\left\{j2\pi l\Delta f \tau\right\},
				\\ 			
				\Lambda_\text{C} (\tau,\nu) & =  T_\text{diff} \sum\nolimits^{L-1}_{l_1=0} \sum\nolimits^{L-1}_{l_2=0 \atop l_2\neq l_1}  A_{l_1}A_{l_2} \exp\left(j\left(\psi_{l_1}-\psi_{l_2}\right)\right) 
				\\ & \ \times  \exp\left\{j2\pi \left[\left(\left(l_1-l_2\right)\Delta f -\nu\right) T_\text{avg} + l_2\Delta f \tau \right]  \right\}
				\\ & \ \times \text{sinc}\left\{ \left[\left(l_1-l_2\right)\Delta f -\nu\right] T_\text{diff} \right\}. 
			\end{aligned}
		\end{equation*}
		Here $T_\text{avg}$ and $T_\text{diff}$ are defined as $T_\text{avg} = \frac{T_\text{max}+T_\text{min}}{2}$ and $T_\text{diff} = T_\text{max}-T_\text{min}$, where $T_\text{min}=\max\left(0,\tau\right)$, and $T_\text{max}=\min\left(T_p,T_p+\tau\right)$. In addition, $\text{sinc}(x) = \frac{\sin(\pi x)}{\pi x}$. 
	\end{proposition}
	\noindent \textit{Proof:} See Appendix \ref{appendix1}. 
	\hfill $\blacksquare$
	\vspace{\baselineskip}

	Evidently, $\Lambda_\text{S}(\tau,\nu)$ represents the superposition of $L$ self-AF components, while $\Lambda_\text{C}(\tau,\nu)$ contains the $L(L-1)$ cross-AF components.
	Next, we aim to derive the expectations and variances of $\Lambda_\text{S}(\tau,\nu)$ and $\Lambda_\text{C}(\tau,\nu)$. To proceed, we first present the following assumption.
	
	\begin{assumption}
		In this article, we consider an ``arbitrary'' constellation shape following the rule that all constellation points are symmetrically distributed with respect to the center, where points in each concentric circle are with the same probabilities. For example, graphical illustration of 16-QAM based constellation format is displayed in Fig. \ref{fig1x}, where $p_1=\cdots=p_4$, $p_5=\cdots=p_{12}$, $p_{13}=\cdots=p_{16}$, and $\sum^{Q=16}\nolimits_{q=1}p_q=1$.
		This rule evidently guarantees that 
		$
		\mathbb{E}\{ A_l \exp(j\psi_l)\} = 0.
		$
	\end{assumption}


	
	For notational convenience, we use $\Lambda$, $\Lambda_S$, and $\Lambda_C$ in the following. The expectation and variance of $\Lambda$ may then be represented as
	\begin{equation}
		\begin{aligned}\label{ep}
			\mathbb{E}[\Lambda] = \mathbb{E}[\Lambda_\text{S}] + \mathbb{E}[\Lambda_\text{C}],
		\end{aligned} 
	\end{equation}
	and
		\begin{align}\label{Eq2}
			& \sigma^2_{\Lambda} =  \mathbb{E}[|\Lambda|^2] - |\mathbb{E}[\Lambda]|^2  = \underbrace{\mathbb{E}[|\Lambda_\text{S}|^2] - |\mathbb{E}[\Lambda_\text{S}]|^2}_{\sigma^2_\text{S}} \\ & + \underbrace{\mathbb{E}\{\left|\Lambda_\text{C}\right|^2\} - \left|\mathbb{E}\{\Lambda_\text{C}\}\right|^2 }_{\sigma^2_\text{C}}
			+2\Re\left\{\mathbb{E}\left(\Lambda_\text{S}\Lambda^*_\text{C}\right)-\mathbb{E}\left(\Lambda_\text{S}\right)\mathbb{E}^*\left(\Lambda_\text{C}\right)\right\}.\nonumber
		\end{align} 

	\begin{proposition}
	We have $\mathbb{E}\left\{\Lambda_\text{C} \right\}=0$, and $\mathbb{E}\left\{\Lambda_\text{S} \Lambda^*_\text{C} \right\}=0$, thereby leading to 
	\begin{equation}
		\begin{aligned}\label{af}
			\mathbb{E}[\Lambda] = \mathbb{E}[\Lambda_\text{S}], \quad 
			\sigma^2_{\Lambda} = 	{\sigma^2_\text{S}} + {\sigma^2_\text{C}}.
		\end{aligned} 
	\end{equation}	
	\end{proposition}

	\noindent \textit{Proof:} See Appendix \ref{appendix2}. 
	\hfill $\blacksquare$
	\vspace{\baselineskip} 		
	
	\subsection{Statistical Characteristics of $\Lambda_\text{S}(\tau,\nu)$}
	Since $\mathbb{E}\{A^2_l\}=1$ holds for arbitrary constellations satisfying the power constraint, we have
	\begin{equation*}
		\begin{aligned}
			\mathbb{E}\{\Lambda_\text{S}\} = T_\text{diff} \text{sinc}\left( -\nu T_\text{diff} \right) \exp\left\{-j2\pi \nu T_\text{avg}\right\}  \\ \times  \sum\nolimits^{L-1}_{l=0} \exp\left\{j2\pi l\Delta f \tau\right\}.
		\end{aligned}
	\end{equation*}
	Evidently, the expectation holds for arbitrary input distribution of constellation symbols, which implies that it does not play a role in the subsequent PCS design.
	Accordingly, we mainly concentrate on the variance of $\Lambda_\text{S}$, which is
	\begin{equation*}
		\begin{aligned}\label{Eq4}
			\sigma^2_\text{S}&  =  \mathbb{E}[\left\vert\Lambda_\text{S}\right\vert^2] - \left\vert \mathbb{E}[\Lambda_\text{S}]\right\vert^2  
			=   T^2_\text{diff} \text{sinc}^2\left( - \nu T_\text{diff} \right) \\ 
			& \quad \times \sum^{L-1}_{l_1=0}\sum^{L-1}_{l_2=0} 	\left\{ \mathbb{E}\left\{A^2_{l_1}A^2_{l_2}\right\} - 1 \right\} 
			\exp\left\{j2\pi (l_1-l_2)\Delta f \tau \right\}.  
		\end{aligned}
	\end{equation*}
	As special cases, for PSK with constant modulus, it is evident that $\sigma^2_\text{S}=0$ due to $\mathbb{E}\left\{A^2_{l_1}A^2_{l_2}\right\}=1$. In contrast, the variance for uniformly/non-uniformly distributed QAM is not zero, since
	\begin{equation}\label{Eq5}
		\mathbb{E}\left\{A^2_{l_1} A^2_{l_2}\right\} =\left\{
		\begin{aligned}
			& { \mathbb{E}\left\{A^4_{l_1}\right\} }, \ & l_1=l_2 \\
			& \mathbb{E}\left\{A^2_{l_1}\right\} \cdot \mathbb{E}\left\{A^2_{l_2}\right\}=1, \ & l_1\neq l_2. 
		\end{aligned}
		\right.
	\end{equation}
	Thanks to (\ref{Eq5}), we may further simplify $\sigma^2_\text{S}$ as
	\begin{equation}\label{Eq6}
		\begin{aligned}
			\sigma^2_\text{S} =  T^2_\text{diff} \text{sinc}^2\left( - \nu T_\text{diff} \right)  \sum\nolimits^{L-1}_{l=0} \left({\mathbb{E}\left\{A^4_{l}\right\} - 1} \right). 
		\end{aligned}
	\end{equation}
	The above result implies that the variances of Doppler sidelobes ($\nu\neq 0$) are very small.
	Therefore, the major impact of randomness is closely related to the zero-Doppler slice known as the autocorrelation function, leading to
	\begin{equation}\label{Eq_zero_Doppler}
		\begin{aligned}
			\sigma^2_\text{S} = LT^2_\text{diff} \left(\mathbb{E}\left\{A^4_{x}\right\} - 1 \right), \ \text{when} \ \nu=0.
		\end{aligned}
	\end{equation}
	Both \eqref{Eq6} and \eqref{Eq_zero_Doppler} indicate that the variance of AF is mainly determined by the fourth moment of the constellation's amplitudes. More precisely, the variance may be reduced by minimizing the fourth moment $\mathbb{E}\left\{A^4_{x}\right\}$. Note that the variance is always non-negative by definition, which demonstrates that the fourth moment of the constellation amplitudes is always greater than or equal to the square of transmit power.
	Hence, we have the following corollary.
	
	\begin{corollary}
		When $\nu=0$, $\sigma^2_\text{S}= 0$ holds only when all constellation symbols have the constant modulus. 
	\end{corollary}
	\noindent \textit{Proof:} Owing to $\sum^{Q}_{q=1}p_qA^2_q=1$ and $\sum^{Q}_{q=1}p_q=1$, we have
	\begin{equation*}
		\begin{aligned}
			\mathbb{E}\left\{A^4_{x}\right\} & =  \mathbb{E}\left\{A^4_{x}\right\} \sum\nolimits^{Q}_{q=1}p_q =   \sum\nolimits^{Q}_{q=1}p_qA^4_q \sum\nolimits^{Q}_{q=1}p_q 
			\\ & \geq  \left( \sum\nolimits^{Q}_{q=1}\sqrt{p_q} A^2_q \sqrt{p_q} \right)^2 
			= \left( \sum\nolimits^{Q}_{q=1}p_qA^2_q \right)^2 
			= 1.
		\end{aligned}
	\end{equation*}
	The equal sign holds when $\frac{\sqrt{p_q} A^2_q}{\sqrt{p_q}}=1$, i.e. $A^2_q=1$, for $\forall q$, leading to the unit modulus of all constellation points, i.e., a ``pseudo'' PSK modulations, as the real PSK has unit modulus and equally spaced phases simultaneously.
	\hfill $\blacksquare$

	\subsection{Statistical Characteristics of $\Lambda_\text{C}$}
	To derive the variance of $\Lambda_\text{C}$, we only need to compute $\mathbb{E}\{|\Lambda_\text{C}|^2\}$, which is formulated in (\ref{Eq7}) at the top of the next page.
	\begin{figure*}
			\begin{align}\label{Eq7}
				\mathbb{E} \left\{|\Lambda_\text{C} (\tau,\nu) |^2\right\}& =   T^2_\text{diff}  \sum^{L-1}_{l_1=0} \sum^{L-1}_{l_2=0,\atop l_2\neq l_1} \sum^{L-1}_{l'_1=0} \sum^{L-1}_{l'_2=0,\atop l'_2\neq l'_1} 
				\mathbb{E}\left\{ A_{l_1}A_{l_2}A_{l'_1}A_{l'_2} 
				\exp\left\{j\left(\psi_{l_1}-\psi_{l_2}-\psi_{l'_1}+\psi_{l'_2}\right)\right\}  \right\} 
				\text{sinc}\left\{ 2\pi\left[\left(l_1-l_2\right)\Delta f -\nu\right] T_\text{diff} \right\}  \nonumber
				\\ &  \times  \text{sinc}\left\{ 2\pi\left[\left(l'_1-l'_2\right)\Delta f -\nu\right] T_\text{diff} \right\} \exp\left\{j2\pi \left[\left(\left(l_1-l_2-l'_1+l'_2\right)\Delta f \right) T_\text{avg} + \left(l_2-l'_2\right)\Delta f \tau \right] \right\}. 
			\end{align}
	\rule{18cm}{1.0pt}
	\end{figure*}
	For further simplifications, one may rely on
	\begin{equation}\label{Eq8}
		\begin{aligned}	
			& \mathbb{E}\left\{A_{l_1}A_{l_2}A_{l'_1}A_{l'_2} \exp\left(j(\psi_{l_1}-\psi_{l_2}-\psi_{l'_1}+\psi_{l'_2})\right) \right\} 
			\\ & = \left\{
			\begin{aligned}
				\mathbb{E}\left\{ A^2_{l_1}A^2_{l'_1} \right\}, \ \ l_1=l_2,l'_1=l'_2 \\
				\mathbb{E}\left\{ A^2_{l_1}A^2_{l_2} \right\}, \ \ l_1=l'_1,l_2=l'_2 \\
				0, \quad \quad \quad \quad \quad \text{otherwise}.
			\end{aligned}
			\right.
		\end{aligned}
	\end{equation}
	Note however that $\Lambda_\text{C}$ is defined when $l_2\neq l_1$ and $l'_2\neq l'_1$ in (\ref{Eq7}). As a consequence, recalling (\ref{Eq8}) suggests that all the non-zero components of $\mathbb{E}\{|\Lambda_\text{C}|^2\}$ are contributed by the constraints of $l_1=l'_1$ and $l_2=l'_2$, yielding
	\begin{equation*}
		\begin{aligned}
			\sigma^2_\text{C} = \mathbb{E}\left\{|\Lambda_\text{C}|^2 \right\} & -   |\mathbb{E}[\Lambda_\text{C}]|^2  = T^2_\text{diff} \sum^{L-1}_{l_1=0} \sum^{L-1}_{l_2=0,\atop {l_2\neq l_1}}  \mathbb{E}\left\{A^2_{l_1}A^2_{l_2} \right\}   \\ &  \times \text{sinc}^2 \left\{ 2\pi\left[(l_1-l_2)\Delta f -\nu\right] T_\text{diff} \right\}. 
		\end{aligned}
	\end{equation*}
	In addition, the condition $l_2\neq l_1$ results in independent random variables $A^2_{l_1}$ and $A^2_{l_2}$. Then exploiting (\ref{Eq5}) yields
	\begin{equation}\label{Eq9}
		\begin{aligned}
			\sigma^2_\text{C} = T^2_\text{diff} \sum\nolimits^{L-1}_{l_1=0} \sum\nolimits^{L-1}_{l_2=0,\atop {l_2\neq l_1}}  \text{sinc}^2 \left\{ 2\pi 
			\left[(l_1-l_2)\Delta f -\nu\right] T_\text{diff} \right\}.
		\end{aligned}
	\end{equation}

	Evidently, when $\nu = 0$, $\sigma^2_\text{C}$ can be approximately omitted owing to $\text{sinc}(2\pi(l_1-l_2)\Delta f)\approx 0$ for $l_1\neq l_2$. In contrast, $\sigma^2_\text{C}$ may be relatively large when $\nu \neq 0$. Nevertheless, regardless of the value of $\sigma^2_\text{C}$, big or small, it is fixed for arbitrary input distribution of constellation symbols, imposing no effect on the subsequent PCS design.
	
	According to (\ref{Eq6}) and (\ref{Eq9}), we denote variances of PSK and QAM as $\sigma^2_\text{S,PSK}$, $\sigma^2_\text{S,QAM}$, $\sigma^2_\text{C,PSK}$ and $\sigma^2_\text{C,QAM}$. Then it is evident that
	\begin{equation}
		\begin{aligned}
			\sigma^2_\text{S,QAM} > \sigma^2_\text{S,PSK} = 0, \quad
			\sigma^2_\text{C,PSK} = \sigma^2_\text{C,QAM}.
		\end{aligned}
	\end{equation}
	This again reflects the superiority of PSK modulation over its QAM counterpart in terms of radar sensing, due to the fact that a smaller variance of AF may effectively lead to improved sensing performance, as the fluctuations in the sidelobes incurred by random data are suppressed. Nevertheless, since QAM generally achieves a better communication performance than that of PSK, one may tradeoff between the AIR and sensing performance of QAM by devising $\sigma^2_\text{S}$ with respect to the input distribution. This paves the way to a PCS approach to balance between S\&C.

	\subsection{AF of OFDM Sequences}
	Next, we extend our previous analysis to OFDM sequences with multiple symbols. In contrast to the AF of a conventional radar pulse train, which is relatively trivial since the same waveform is applied for all pulses, the similar extension for OFDM communication signals becomes more challenging owing to the sequence randomness in multiple symbols. 
	
	With the accumulation of $N$ symbols in the time domain, a train of OFDM sequences is expressed as
   \begin{equation}
		\begin{aligned}
			\tilde{s}(t) = & \sum\nolimits^{N-1}_{n=0} \sum\nolimits^{L-1}_{l=0}A_{n,l}   \exp\left(j\psi_{n,l}\right) \\ & \times \exp\left(j2\pi l \Delta f(t-nT_p)\right) 
			\text{rect}\left( ({t-nT_p})/{T_p} \right),
		\end{aligned}
    \end{equation}
	where $\exp\left(j2\pi l \Delta f(t-nT_p)\right)\triangleq\phi_l(t-nT_p)$.
	According to the definition in \cite{sen2009adaptive,tigrek2012ofdm}, the AF of $\tilde{s}(t)$ is formulated as
		\begin{align}\label{Eq10}
			& \tilde{\Lambda}(\tau,\nu) = \int^{\infty}_{-\infty} \tilde{s}(t)\tilde{s}^*(t-\tau) \exp\left(-j2\pi\nu t\right)dt \nonumber
			\\ & = \sum^{N-1}_{n_1=0}\sum^{N-1}_{n_2=0} \sum^{L-1}_{l_1=0}\sum^{L-1}_{l_2=0} A_{n_1,l_1}A_{n_2,l_2}\exp\left(j(\psi_{n_1,l_1}-\psi_{n_2,l_2})\right) \nonumber
			\\ &  \quad \times \int^{\infty}_{-\infty} \phi_{l_1}(t')\phi^*_{l_2}(t'+n_1T_p-n_2T_p-\tau) \nonumber
			\\ &  \qquad \quad \times \exp\left(-j2\pi\nu (t'+n_1T_p)\right) dt',
		\end{align}
	where $t'= t-n_1T_p$. For notational convenience, we denote 
	\begin{equation*}
		\begin{aligned}
			\tilde{T}_\text{min} = & \max\left(0,\tau+\left(n_2-n_1\right)T_p\right), \\
			\tilde{T}_\text{max} = & \min\left(T_p,T_p+\tau+\left(n_2-n_1\right)T_p\right), \\
			\tilde{T}_\text{avg} = & \left(\tilde{T}_\text{max}+\tilde{T}_\text{min}\right)/2, \quad
			\tilde{T}_\text{diff} = \tilde{T}_\text{max}-\tilde{T}_\text{min}.
		\end{aligned}
	\end{equation*}
	Then the integral in (\ref{Eq10}) can be computed as in (\ref{Eq11}) at the top of this page.
	\begin{figure*}[t]
		\begin{equation}\label{Eq11}
			\begin{aligned}
				& \int^{\infty}_{-\infty} \phi_{l_1}\left(t'\right)\phi^*_{l_2} \left(t'+n_1T_p-n_2T_p-\tau\right)  \exp\left(-j2\pi\nu \left(t'+n_1T_p\right)\right) dt' 
				\\ & 
				=  \exp \left\{ j2\pi \left[l_2\Delta f \left(\left(n_2-n_1\right)T_p+\tau \right) - n_1\nu T_p \right] \right\}
				\int^{\tilde{T}_\text{max}}_{\tilde{T}_\text{min}} \exp\left\{j2\pi \left[\left(l_1-l_2\right)\Delta f -\nu\right] t'\right\} dt'
				\\ & = \tilde{T}_\text{diff} \text{sinc} \left\{ \left[ \left(l_2-l_1\right)\Delta f - \nu \right]  \tilde{T}_\text{diff} \right\} \exp \left\{ j2\pi \left[ \left(\left(l_1-l_2\right)\Delta f -\nu\right) \tilde{T}_\text{avg} + l_2\Delta f \left(\left(n_2-n_1\right)T_p+\tau\right) - n_1\nu T_p \right] \right\}.
			\end{aligned}
		\end{equation}
	\rule{18cm}{1.0pt}
	\end{figure*}

	In the same manner, we consider the self and cross components of AF, expressed as $\tilde{\Lambda} = \tilde{\Lambda}_\text{S} +\tilde{\Lambda}_\text{C}$, where $\tilde{\Lambda}_\text{S}$ is contributed by the $n_1=n_2$ and $l_1=l_2$ terms, i.e.,
	\begin{equation}
		\begin{aligned}\label{Eq12}
			\tilde{\Lambda}_\text{S} & =  T_\text{diff} \text{sinc}\left( -\nu T_\text{diff} \right) \exp\left\{-j2\pi \nu T_\text{avg}\right\} \sum^{N-1}_{n=0} \sum^{L-1}_{l=0} A^2_{n,l} \\ & \times \exp\left\{j2\pi \left( l\Delta f \tau - n\nu T_p\right)\right\},
		\end{aligned}
	\end{equation}
	and $\tilde{\Lambda}_\text{C}$ is contributed by all remaining components, which is formulated in (\ref{Eq13}). Note that in (\ref{Eq12}) $T_\text{diff}$ and $T_\text{avg}$ are adopted rather than $\tilde{T}_\text{diff}$ and $\tilde{T}_\text{avg}$, due to $n_1=n_2$.
	
	\begin{figure*}[t]
			\begin{align}\label{Eq13}
				\tilde{\Lambda}_\text{C} (\tau,\nu) = & \sum\nolimits^{N-1}_{n=0} \sum\nolimits^{L-1}_{l_1=0} \sum\nolimits^{L-1}_{l_2=0 \atop l_2\neq l_1}
				A_{n,l_1} A_{n,l_2} \exp\left(j \left(\psi_{n,l_1}-\psi_{n,l_2} \right)\right) \nonumber
				\\ &  \times  \underbrace{ T_\text{diff} \text{sinc} \left\{ \left[\left(l_2-l_1\right)\Delta f - \nu \right]  T_\text{diff} \right\} 
					\exp \left\{ j2\pi \left[ \left(\left(l_1-l_2\right)\Delta f -\nu\right) T_\text{avg} + l_2\Delta f \tau - n\nu T_p \right] \right\} }_{R_1(n,l_1,l_2)}\nonumber
				\\ 
				& + \sum\nolimits^{N-1}_{n_1=0} \sum\nolimits^{N-1}_{n_2=0 \atop n_2 \neq n_1} \sum\nolimits^{L-1}_{l_1=0} \sum\nolimits^{L-1}_{l_2=0}
				A_{n_1,l_1} A_{n_2,l_2} \exp\left(j\left(\psi_{n_1,l_1}-\psi_{n_2,l_2}\right)\right) 
				\\ &  \times  \underbrace{ \tilde{T}_\text{diff} \text{sinc} \left\{ \left[\left(l_2-l_1\right)\Delta f - \nu \right]  \tilde{T}_\text{diff} \right\} 
					\exp \left\{ j2\pi \left[ \left(\left(l_1-l_2\right)\Delta f -\nu\right) \tilde{T}_\text{avg} + l_2\Delta f \left(\left(n_2-n_1\right)T_p+\tau\right) - n_1\nu T_p \right] \right\} }_{R_2(n_1,n_2,l_1,l_2)}. \nonumber
			\end{align}
	\rule{18cm}{1.0pt}
	\end{figure*}

	Similar to the case of ${\Lambda}_\text{S}$ and ${\Lambda}_\text{C}$, the statistical characteristics of $\tilde{\Lambda}_\text{S}$ and $\tilde{\Lambda}_\text{C}$ are summarized below.
	
	\begin{proposition}
		If the same constellation modulation is adopted for all OFDM symbols, the variance of $\tilde{\Lambda}_\text{S}$, denoted as $\tilde{\sigma}^2_\text{S}$, is $N$ times the accumulation of that of a single symbol, i.e., 
		\begin{equation}
			\begin{aligned}
				\tilde{\sigma}^2_\text{S} = N \sigma^2_\text{S}.
			\end{aligned}
		\end{equation}
	\end{proposition}
	
	\noindent \textit{Proof:} See Appendix \ref{appendix3}. 
	\hfill $\blacksquare$
	
	\begin{proposition}
		The variance of $\tilde{\Lambda}_\text{C}$, denoted as $\tilde{\sigma}^2_\text{C}$, does not affect the constellation shaping, since it is fixed for arbitrary input distribution of constellation symbols, i.e.,
		\begin{equation}
			\begin{aligned}
				\tilde{\sigma}^2_\text{C} = \text{constant}, \ \forall p_q, q\in \{1,2,\cdots,Q\}.
			\end{aligned}
		\end{equation}
	\end{proposition}
		
	\noindent \textit{Proof:} See Appendix \ref{appendix4}. 
	\hfill $\blacksquare$
	\vspace{\baselineskip} 		
	
	Since the value of $\tilde{\sigma}^2_\text{C}$ (or equivalently, $\sigma^2_\text{C}$) is constant for arbitrary input distribution of constellation symbols, we are only able to control $\tilde{\sigma}^2_\text{S}$ (or equivalently, $\sigma^2_\text{S}$), by optimizing the input distribution through PCS. In this manner, the randomness of OFDM signaling can be scaled, thereby affecting the S\&C performance simultaneously.

	\section{Communication Performance Evaluation}\label{sec3}
	We adopt the AIR in an additive white Gaussian noise (AWGN) channel as a communication performance indicator. The received communication signal is expressed in a simple linear form as $y_c(t)=s(t)+n_c(t)$. After the discretization, the discrete signal in the time domain is recast in the matrix-vector form as  $\mathbf{y}_c=\mathbf{D}^H\mathbf{x}+\mathbf{n}_c$, where $\mathbf{y}_c=[y_c(0),\cdots,y_c(L-1)]^T$, $\mathbf{x}=[A_0e^{j\psi_0},\cdots,A_{L-1}e^{j\psi_{L-1}}]^T$,  $\mathbf{n}_c=[n_c(0),\cdots,n_c(L-1)]^T$, and $\mathbf{D}$ is the discrete Fourier transform (DFT) matrix.
	By transforming $\mathbf{y}_c$ into the frequency domain, we have
	\begin{equation}
		\begin{aligned}\label{dft}
			\mathbf{y} = \mathbf{x} + \mathbf{n},
		\end{aligned}
	\end{equation}	
	where $\mathbf{y}=\mathbf{D}\mathbf{y}_c$, and $\mathbf{n}=\mathbf{D}\mathbf{n}_c$. Here, $\mathbf{n}\sim \mathcal{CN}(\mathbf{0},\sigma^2\mathbf{I}_L)$.

	In point-to-point channels, the AIR of $L$ parallel subchannels is characterized by the input-output mutual information (MI), which is \cite{carmon2015comparison}
	\begin{equation}
		\begin{aligned}
			I\left(\mathbf{x};\mathbf{y}\right) = \sum^{L-1}\nolimits_{l=0} I\left(x_l;y_l\right),
		\end{aligned}
	\end{equation}	
	where $x_l$ and $y_l$ represents the $l$th subchannel elements of $\mathbf{x}$ and $\mathbf{y}$, respectively, and $I\left(x_l;y_l\right)$ denotes the MI in the $l$th subchannel. It is not surprising to see that $I\left(x_l;y_l\right)$ are equal for all subchannels, due to the fact that the same constellation codebook is adopted at all subcarriers. This suggests that we only need to analyze the MI of a single subchannel, since the overall AIR may be computed as a summation of the rates over $L$ sub-channels, namely, $I\left(\mathbf{x};\mathbf{y}\right) = L I\left(x;y\right)$, where the subscripts $l$ is omitted for notational convenience. 
		
	The MI $I\left(x,y\right)$ is derived as
	\begin{equation}
		\begin{aligned}\label{MInf}
			I\left(x;y\right) = \mathbb{E}_{X,Y}\left\{\log\frac{p(y|x)}{{p(y)}} \right\} 
			= H_Y(y)-H_{Y|X}(y|x),
		\end{aligned}
	\end{equation}
	where 
	\begin{equation*}
		\begin{aligned}
			H_{Y|X}(y|x) & = \log(\pi e\sigma^2), \ \text{and} \\
			H_Y(y) & = - \sum\nolimits_{y} \left[ \log \sum\nolimits_{x'} p(y|x') p(x') \right] \sum\nolimits_{x} p(y|x)p(x)
		\end{aligned}
	\end{equation*}
	denote the conditional entropy of $y$ given $x$, and the entropy of $y$, respectively. Despite that constellation points are discrete, the received signal $y$ is a continuous random variable. For the sake of technical convenience, $y$ is discretized for the follow-up processing. As a result, we use discrete sum rather than continuous integral in calculating $H_Y(y)$. 
 
    Since $p(y) = \sum_{x} p(y|x)p(x)$ is the sum of Gaussian probability density functions (PDFs) $p(y|x)$ weighted by the input distribution $p(x)$, the closed-form of $H_Y(y)$ is generally non-obtainable due to Gaussian mixture (GM) distributed $y$ \cite{gu2017information}. To observe the AIR performance, we therefore approximately compute $H(y)$ using Monte Carlo numerical integrals:
	\begin{equation}\label{Eq15}
		\begin{aligned}
			H_Y(y) = & -\mathbb{E}_Y \left[ \log \sum\nolimits_{x } p(y|x) p(x) \right] \\
			\approx & -\frac{1}{N_\text{MC}}\sum\nolimits^{{N_\text{MC}}}_{m=1} \log \sum\nolimits_{x } p(y_m|x) p(x),
		\end{aligned}
	\end{equation}
	where ${N_\text{MC}}$ represents the number of Monte Carlo trials, $y_m$ denotes the $m$th observation and its conditional PDF $p(y_m|x)$ is with standard Gaussian forms in the $k$th trial, for each $x$ in the given constellation. By doing so, the entropy $H_Y(y)$ can be accurately approximated when ${N_\text{MC}}$ is sufficiently large, thereby obtaining the accurate value of AIR.

	\section{PCS Approaches for ISAC}\label{sec5}
	\subsection{Optimal PCS Optimization Modeling}\label{sec5.1}
	
    On the basis of S\&C metrics, i.e., the fourth-moment of constellations and the AIR, both of which are determined by the input distribution, we hereby present an optimal PCS model for ISAC that maximizes the MI, under the sensing, power and probability constraints:
	
	\begin{equation*}\label{Eq16}
		\left({\mathcal{P}1} \right)
		\left\{
		\begin{aligned}
			\max_{\mathbf{p}}  & \ I\left(\mathbf{p}\right)  \triangleq I\left(x;y\right)
			\\
			s.t. \ & C_1: \sum\nolimits_{x}p(x)A^4_x = c_0, \quad
			C_2: \sum\nolimits_{x}p(x)A^2_x = 1 \\
			& C_3: \sum\nolimits_{x}p(x) = 1, \quad
			C_4: p(x) \geq 0, \ \forall x \in \mathcal{X}
		\end{aligned}
		\right.
	\end{equation*}
	where $\mathbf{p} = [p_1, p_2,..., p_Q]^T$ represents the discrete input probability distribution vector of a given constellation. Here, $c_0$ is a desired value of the fourth-moment of constellation amplitudes, which can be scaled by optimizing $\mathbf{p}$, in order to control the variation of the sidelobes of AF for sensing.
    By recalling (\ref{MInf}), the MI in $\mathcal{P}1$ is recast as 
	\begin{equation*}
		\begin{aligned}
			\mathbb{E}_{X,Y} & \left\{\log\frac{p(y|x)}{{p(y)}} \right\} = \sum\nolimits_{x} \sum\nolimits_{y} p(x)p(y|x)\log\frac{p(y|x)}{p(y)}
			\\ & = \max\nolimits_{\mathbf{q}} \ \underbrace{\sum\nolimits_{x} \sum\nolimits_{y} p(x)p(y|x)\log\frac{q(x|y)}{p(x)}}_{\mathcal{F}\left(\mathbf{p},\mathbf{q}\right)},
		\end{aligned}
	\end{equation*}
	where $\mathbf{q}$ denotes a transition matrix from $\mathcal{Y}$ to $\mathcal{X}$ \cite{yeung2008information}, with $\mathcal{X}$ and $\mathcal{Y}$ representing input and output alphabets of $x$ and $y$, respectively. We further reformulate $\mathcal{P}1$ as 
	\begin{equation*}
		\left({\mathcal{P}2} \right)
		\left\{
		\begin{aligned}
			\max_{\mathbf{p}} & \max_{\mathbf{q}} \  \mathcal{F}\left(\mathbf{p},\mathbf{q}\right), \\ s.t. \ & C_1\sim C_4.
		\end{aligned}
		\right.
	\end{equation*}
	The model $\mathcal{P}2$ is a convex optimization problem, since it aims to maximize an objective function jointly concave in $\mathbf{p}$ and $\mathbf{q}$, under linear constraints $C_1$-$C_4$. It is worth highlighting that the classic MI maximization problem can be readily solved by the standard Blahut-Arimoto algorithm \cite{yeung2008information} if the sensing constraint is not imposed. Nevertheless, by incorporating $C_1$ and $C_2$, $\mathcal{P}2$ becomes more challenging, and the conventional methods are not straightforwardly applicable. To that end, we elaborate on an MBA algorithm tailored for $\mathcal{P}2$ by relying on alternating optimization between $\mathbf{p}$ and $\mathbf{q}$.

	As evidenced in \cite[Lemma 9.1]{yeung2008information}, given the input distribution $\mathbf{p}$, the optimal $\mathbf{q}$ that maximizes the MI is expressed as
	\begin{equation}
		\begin{aligned}\label{MI}
			q(x|y) = \frac{p(x)p(y|x)}{\sum_{x'}p(x')p(y|x')},
		\end{aligned}
	\end{equation} 
	which can be proved by applying the divergence inequality.
	In order to solve for $\mathbf{p}$ that maximizes the objective under a given $\mathbf{q}$ (i.e., maximizing the MI under constraints $C_1\sim C_4$ for obtaining the AIR, namely, the constrained capacity), we now resort to the Lagrange multiplier to seek for the optimal $\mathbf{p}$ by temporarily ignoring the constraint $C_4$, and thus obtain
	\begin{align}
		\mathcal{L} = & \sum_x \sum_y p(x)p(y|x)\log\frac{q(x|y)}{p(x)} - \lambda_1 \left(\sum_x p(x)A^4_x-c_0\right) \nonumber
		\\ & - \lambda_2 \left(\sum\nolimits_x p(x)A^2_x-1\right) - \lambda_3 \left(\sum\nolimits_x p(x)-1\right).
	\end{align}
	For the sake of convenience, the natural logarithm is assumed. Differentiating with respect to $p(x)$ yields
	\begin{equation}
		\begin{aligned}
			\frac{\partial\mathcal{L}}{\partial p(x)} = & \sum\nolimits_y p(y|x) \log q(x|y) - \log p(x) 
			\\ & - \lambda_1 A^4_x - \lambda_2 A^2_x 
			- \lambda_3 -1.
		\end{aligned}
	\end{equation}
	By setting $\frac{\partial\mathcal{L}}{\partial p(x)} = 0$, we have
	\begin{equation}
		\begin{aligned}
			p(x) = & \exp\left\{ \sum\nolimits_y p(y|x) \log q(x|y) - \lambda_1 A^4_x - \lambda_2 A^2_x \right\} \\ & \times
			\exp\left\{ - \lambda_3 -1  \right\}.
		\end{aligned}
	\end{equation}
	Recalling $C_3$, we further simplify the result of $p(x)$ as follows:
	\begin{equation*}
		\begin{aligned}
			p(x) =  \frac{\exp\left\{\sum_y p(y|x)\log q(x|y) -\lambda_1A^4_x -\lambda_2A^2_x \right\} } {\sum_{x}\exp\left\{\sum_yp(y|x)\log q(x|y) -\lambda_1A^4_{x} -\lambda_2A^2_{x} \right\}}.
		\end{aligned}
	\end{equation*}
	It is immediately observed that $p(x)$ satisfies $C_4$ due to the fact that exponentials are positive.
	
	Given $\lambda_1$ and $\lambda_2$, we may now define the $k$th iterative probabilities $q^{(k)}(x|y)$ and $p^{(k+1)}(x)$ for $k\geq 0$ as
		\begin{align}
			q^{(k)}(x|y) = & \frac{p^{(k)}(x)p(y|x)}{\sum_{x'}p^{(k)}(x')p(y|x')}, \label{iteration_formats}\\
			p^{(k+1)}(x) = & \frac{e^{\sum_y p(y|x)\log q^{(k)}(x|y) -\lambda_1A^4_x -\lambda_2A^2_x } } {\sum_{x}e^{\sum_yp(y|x)\log q^{(k)}(x|y) -\lambda_1A^4_{x} -\lambda_2A^2_{x} }}.
		\end{align}
    Then we have
	\begin{equation*}
		\begin{aligned}
			& \mathcal{F}\left(\mathbf{p}^{(k+1)},\mathbf{q}^{(k+1)}\right)=  \max\nolimits_{\mathbf{p}}\ \mathcal{F}\left(\mathbf{p},\mathbf{q}^{(k+1)}\right) \\ & \geq \mathcal{F}\left(\mathbf{p}^{(k)},\mathbf{q}^{(k+1)}\right) =  \max\nolimits_{\mathbf{q}}\ \mathcal{F}\left(\mathbf{p}^{(k)},\mathbf{q}\right) \geq
			\mathcal{F}\left(\mathbf{p}^{(k)},\mathbf{q}^{(k)}\right),
		\end{aligned}
	\end{equation*}
	which implies that the alternating optimization converges and admits a globally optimal solution, since the concave objective function of $\mathcal{P}_2$ is bounded from the above by the entropy of the constellation. We refer readers to \cite{yeung2008information} for a detailed convergence proof of the BA algorithm.
	
	Substituting \eqref{iteration_formats} into $C_1$ and $C_2$ yields
	\begin{equation}\label{Eq19}
		\begin{aligned}
			& f_1(\lambda_1,\lambda_2) = \sum\nolimits_x (A^2_x-1) g(\lambda_1,\lambda_2) = 0,  \\
			& f_2(\lambda_1,\lambda_2) = \sum\nolimits_x (A^4_x-c_0) g(\lambda_1,\lambda_2) = 0, 
		\end{aligned}
	\end{equation}
	where
	\begin{align}\label{Eq20}
		g(\lambda_1,\lambda_2) = \exp\left\{\sum\nolimits_y p(y|x)\log q^{(k)}(x|y) -\lambda_1A^4_x -\lambda_2A^2_x \right\}.
	\end{align}
	Note however that the discrete integral in (\ref{Eq20}) is defined with respect to the GM random variable $Y$. Therefore, it should also be computed with Monte Carlo integration, expressed as
		\begin{align}\label{MC_integral}
			\sum\nolimits_y p & (y|x) \log q^{(k)}(x|y) = \sum\nolimits_y \frac{p(y|x)}{p(y)}\log q^{(k)}(x|y) p(y) \nonumber
			\\
			& = \sum\nolimits_y \frac{p(y|x)}{\sum_x p(y|x)p(x)}\log q^{(k)}(x|y) p(y) 
			\\
			& \approx \frac{1}{N_\text{MC}} \sum\nolimits^{N_\text{MC}}_m \frac{p(y_m|x)}{\sum_x p(y_m|x)p^{(k)}(x)}\log q^{(k)}(x|y_m), \nonumber
		\end{align}
	where $p(y_m|x) = \exp\left(-|y_m-x|^2/\sigma^2\right)/(\pi\sigma^2)$ for $x\in \mathcal{X}$.
	
	To proceed, the next step involves determining Lagrange multipliers $\lambda_1$ and $\lambda_2$ in (\ref{Eq19}). While a brute-force 2D grid search may be effective in achieving this goal, it leads to an unbearable computational burden. Towards that aim, we develop a tailored Newton's method for this problem. It is of significant importance to highlight the sensibility of Newton's method on the choice of initial values. In order to acquire a reliable initial value that balances between accuracy and computational complexity, we employ a relatively coarse 2D grid search for the initialization of the Newton's method, i.e.,
	\begin{equation}
		\begin{aligned}\label{Eq21}
			\left< \lambda^{(0)}_1,\lambda^{(0)}_2 \right> = \arg \min_{\left<\lambda_1,\lambda_2\right>} \left\Vert 
			\left[f_1\left( \lambda_1,\lambda_2 \right);
			f_2\left( \lambda_1,\lambda_2 \right) \right]
			\right\Vert.
		\end{aligned}
	\end{equation}
	The solver is completed when the objective converges to an extremely small positive value, thus attaining a relatively accurate initialization. 
	Then the Lagrange multiplier vector $\bm{\lambda}=[\lambda_1,\lambda_2]^T$ may be updated with Newton's method, where the $\ell$th iteration of  $\bm{\lambda}$ is updated as \cite{burden2011numerical}
	\begin{equation}
		\begin{aligned}\label{Eq22}
			\bm{\lambda}^{(\ell+1)} = \bm{\lambda}^{(\ell)} - \left[ \mathbf{F}'(\bm{\lambda}^{(\ell)}) \right]^{-1} \mathbf{F}\left(\bm{\lambda}^{(\ell)}\right).
		\end{aligned}
	\end{equation}
	In (\ref{Eq22}), $\mathbf{F}\left(\bm{\lambda}^{(\ell)}\right) = \left[f_1\left(\lambda^{(\ell)}_1,\lambda^{(\ell)}_2\right),f_2\left(\lambda^{(\ell)}_1,\lambda^{(\ell)}_2\right)\right]^T$, and $\mathbf{F}'\left(\bm{\lambda}^{(\ell)}\right)$ is the corresponding Jacobian matrix.
	
	The two iterative processes above are terminated when the difference between adjacent two iterations is lower than the convergence tolerance $\varepsilon$, e.g., $\varepsilon=10^{-5}$. For clarity, the main steps of MBA algorithm for solving the optimal PCS approach are summarized in the Algorithm 1.

	\begin{algorithm}[!t]
		\caption{MBA algorithm for $\mathcal{P}2$.}\label{alg:alg1}
		\renewcommand{\algorithmicrequire}{\textbf{Input:}}
		\renewcommand{\algorithmicensure}{\textbf{Output:}}
		\begin{algorithmic}[1] 
			\REQUIRE initialization of $\mathbf{p}^{(0)} = 1/Q\cdot \mathbf{1}_Q$, constellation amplitudes $A_x$,
			convergence tolerance $\varepsilon$; \\ 
			\ENSURE desired constellation probabilities of $\mathbf{p}$.
			\FOR {$k=0,1,\cdots$}
			\STATE $q^{(k)}(x|y) \leftarrow \frac{p^{(k)}(x)p(y|x)}{\sum_{x'}p^{(k)}(x')p(y|x')}$;
			\STATE Initialization of Newton's method: estimate with (\ref{Eq21});
			\FOR {$\ell=0,1,2,\cdots$}
			\STATE Update the Lagrange multiplier vector with (\ref{Eq22});
			\IF {$\left\Vert \bm{\lambda}^{(\ell+1)}-\bm{\lambda}^{(\ell)} \right\Vert^2 \leq \varepsilon$ }
			\STATE Record $\ell_0 = \ell+1$;		
			Break;
			\ENDIF
			\ENDFOR
			\STATE Substitute (\ref{MC_integral}) and $\bm{\lambda}^{(\ell_0)}$ into (\ref{Eq20}):
			\begin{equation*}
				\begin{aligned}
					g\left(\lambda^{(\ell_0)}_1,\lambda^{(\ell_0)}_2 \right) &  \leftarrow \exp\Bigg\{\frac{1}{{N_\text{MC}}} \sum^{{N_\text{MC}}}_m \frac{p(y_m|x)}{\sum_x p(y_m|x)p^{(k)}(x)} \\ & \times \log q^{(k)}(x|y_m) -\lambda^{(\ell_0)}_1A^4_x -\lambda^{(\ell_0)}_2A^2_x \big\}.
				\end{aligned}
			\end{equation*}
			\STATE Update $p(x)$: $p^{(k+1)}(x) \leftarrow \frac{ g\left(\lambda^{(\ell_0)}_1,\lambda^{(\ell_0)}_2 \right) } {\sum_{x} g\left(\lambda^{(\ell_0)}_1,\lambda^{(\ell_0)}_2 \right) }$;
			\IF {$\left\Vert \mathbf{p}^{(k+1)}-\mathbf{p}^{(k)} \right\Vert^2 \leq \varepsilon$ }
			\STATE Record $k_0 = k+1$;
			Break;
			\ENDIF		
			\ENDFOR
			\RETURN the latest estimate of $\mathbf{p}^{(k_0)}$.
		\end{algorithmic}
		\label{alg1}
	\end{algorithm}

	\subsection{Heuristic PCS: A Low Complexity Approach}\label{sec4}

	One may observe that solving $\mathcal{P}2$ requires quantizing the continuous random variable $y$, in an effort to numerically evaluate \eqref{MC_integral} in every iteration by Monte Carlo integration, which incurs significant computational overheads. To reduce the complexity of the PCS design, we propose a heuristic approach without explicitly involving $y$, which is
 
	\begin{equation}\label{Eq14}
		\left({\mathcal{P}3}\right)
		\left\{
		\begin{aligned}
			\min_{\mathbf{p}} & \  \left|\sum\nolimits_{x}p(x)A^4_x - c_0 \right|^2
			\\
			s.t. \ & C_2, C_3, C_4.
		\end{aligned}
		\right.
	\end{equation}
	Again, $c_0$ is a preset value aiming to control the fourth moment of the constellation, thereby adjusting the variance of AF. We use the term ``heuristic'' on account of the absence of AIR in $\mathcal{P}3$, despite that the communication performance may still be \textit{passively} scaled, due to the variation of prior probabilities of the constellation through controlling its fourth moment. Intuitively, since the minimal fourth moment $\mathbb{E}\left\{A^4_{x}\right\} = 1$ is attainable for PSK, by varying $c_0$ from $1$ to the value of a uniform QAM constellation, one may also expect that the communication performance can be scaled from the PSK's AIR to that of QAM. Indeed, such a tradeoff will be observed in our simulation results in Sec. \ref{sec6}.
	
    Note that $\mathcal{P}3$ is a convex quadratic optimization problem with linear constraints, which can be readily solved by various well-established algorithms. A more strict formulation would be to directly enforce the equality constraint $C_1$ instead of adopting the least-squares objective function in $\mathcal{P}3$, in which case the constraints $C_1$, $C_2$ and $C_3$ constitute a linear equation system with respect to the non-negative vector $\mathbf{p}$.
	As demonstrated earlier, the probabilities of constellation points within each concentric circle are equal. Therefore, the computational dimensionality of this linear equation system may be significantly reduced. In this spirit, $C_1$, $C_2$ and $C_3$ can be reconfigured into a compact form, i.e.,
	\begin{equation}
		\begin{aligned}\label{Eq17}
			\underbrace{
				\left( \begin{array}{cccc}
					A^4_1 & A^4_2 & \cdots & A^4_W, \\
					A^2_1 & A^2_2 & \cdots & A^2_W, \\
					1     & 1     & \cdots & 1
				\end{array}\right) }_{\mathbf{A}_W}
			\left( \begin{array}{c}
				\bar{p}_1  \\
				\bar{p}_2 \\
				\vdots  \\
				\bar{p}_W
			\end{array}\right) 
			=
			\left( \begin{array}{c}
				c_0  \\
				1 \\
				1 
			\end{array}\right), \forall \bar{p}_w \ge 0,				
		\end{aligned}
	\end{equation}
	where $W$ denotes the number of concentric circles, and $\bar{p}_w$ represents the identical probability in the $w$th circle.
	
	Given a 16-QAM constellation, the presence of three concentric circles ($W=3$) simplifies the probability calculation, as it only necessitates three probability values. Furthermore, given that the matrix $\mathbf{A}_W$ is now full-rank, (\ref{Eq17}) admits a unique solution while satisfying constraint $C_4$. Importantly, it is noteworthy that this solution is independent of the AIR, implying that solving both $\mathcal{P}2$ and $\mathcal{P}3$ yields identical results for the 16-QAM constellation. In contrast, for a 64-QAM ($W=9$) constellation, (\ref{Eq17}) becomes under-determined. In this case, there may exist multiple feasible solutions, indicating that the solution of $\mathcal{P}3$ is merely one among those of $\mathcal{P}2$. Importantly, this solution does not necessarily represent the optimal one that maximizes the AIR, underscoring that $\mathcal{P}2$ is indeed an optimal PCS model compared to $\mathcal{P}3$. This also highlights the need of specialized algorithms for higher-order QAM modulation schemes, i.e., seeking for a solution of the linear equation system \eqref{Eq17} via the convex programming \eqref{Eq14}.
	
	Finally, we highlight that both $\mathcal{P}2$ and $\mathcal{P}3$ may be implemented in an \textit{offline} manner, where the one-to-one mapping between each $\mathbf{p}$ and each value of $c_0$ may be computed and stored in a look-up table for practical use.

	\section{Simulations}\label{sec6}
	We consider OFDM signals with normalized bandwidth and $L=64$ subcarriers. 
    Conventional radar AF performance is assessed by evaluating $10\log_{10}\vert{\Lambda}(\tau,\nu)\vert^2$ due to the deterministic $\Lambda(\tau,\nu)$. While for random OFDM signaling, all the AFs are evaluated in accordance with their average AF performance, i.e., 
	\begin{equation}
		\begin{aligned}
			\bar{\Lambda}(\tau,\nu) = & 10\log_{10} \frac{1}{{N_\text{MC}}}\sum\nolimits^{N_\text{MC}}_{m=1}|\Lambda_m(\tau,\nu)|^2,
		\end{aligned}
	\end{equation} 
	where ${N_\text{MC}}=5000$ trials are used, and $\Lambda_m(\tau,\nu)$ represents the $m$th realization of the random AF. The average AFs are normalized as well. 
	Recalling (\ref{ep}), (\ref{Eq2}) and (\ref{af}), we observe that
	\begin{equation}
	\begin{aligned}
		\bar{\Lambda}(\tau,\nu) \approx & 10\log_{10} \mathbb{E}\left\{|\Lambda|^2\right\} 
		=  10\log_{10} \left(\sigma^2_{\Lambda} + |\mathbb{E}[\Lambda]|^2\right)
		\\ = & 
		10\log_{10} \left({\sigma^2_\text{S}} + {\sigma^2_\text{C}} + |\mathbb{E}[\Lambda_S]|^2\right).
	\end{aligned}
	\end{equation} 
	Then it is evident that the sidelobes of AF are only determined by the input distribution vector $\mathbf{p}$ through $\sigma^2_\text{S}$, since both $\sigma^2_\text{C}$ and $\mathbb{E}[\Lambda_S]$ are irrelevant to 
	$\mathbf{p}$.
	
	\subsection{AF Performance}
\begin{figure}[!t]
		\centering
		\subfigure[AF $\bar{\Lambda}(\tau,\nu)$] { \label{fig1a}
			\includegraphics[width=1.45in]{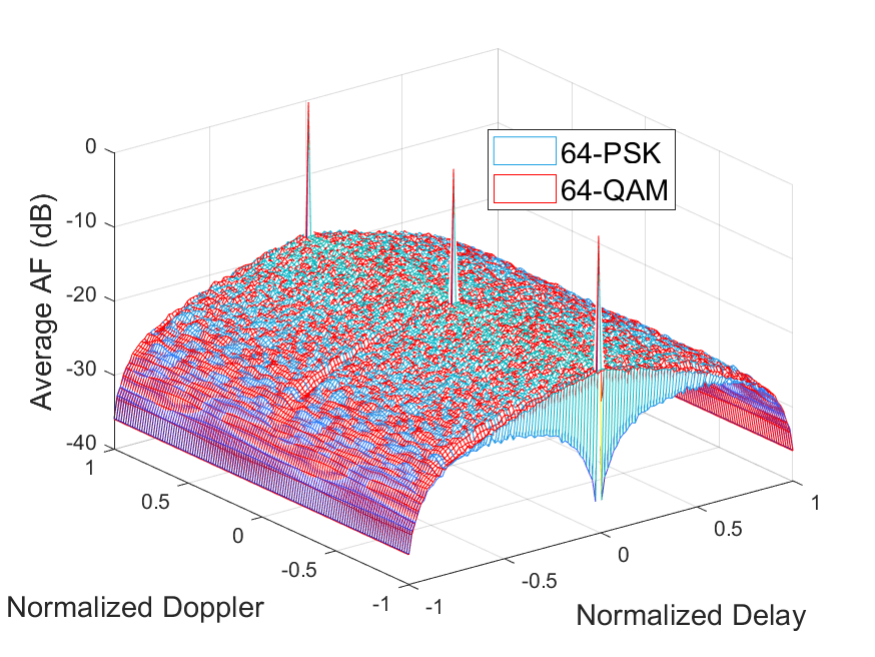}
		}
		\subfigure[Zero Doppler slice $\bar{\Lambda}(\tau,0)$] { \label{fig1b}
			\includegraphics[width=1.45in]{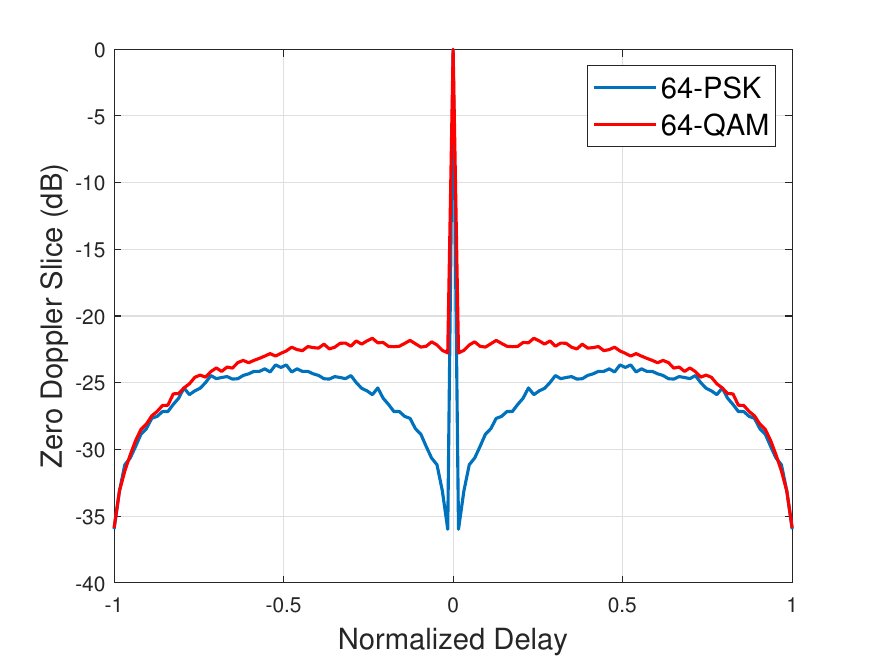}
		}
		\subfigure[Single realization of zero Doppler slice for 64-PSK.] { \label{fig1c}
			\includegraphics[width=1.45in]{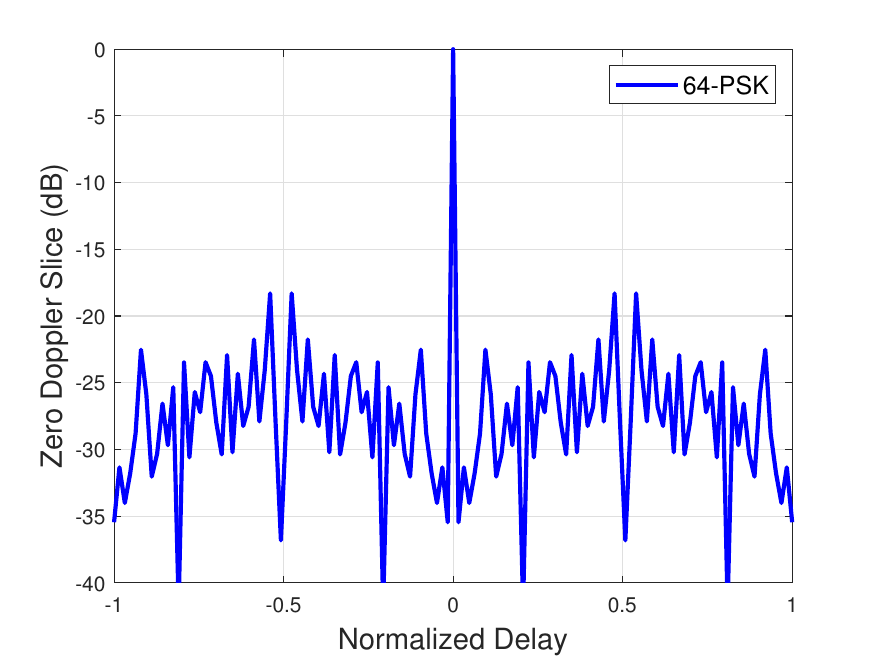}
		}
		\subfigure[Single realization of zero Doppler slice for 64-QAM.] { \label{fig1d}
			\includegraphics[width=1.45in]{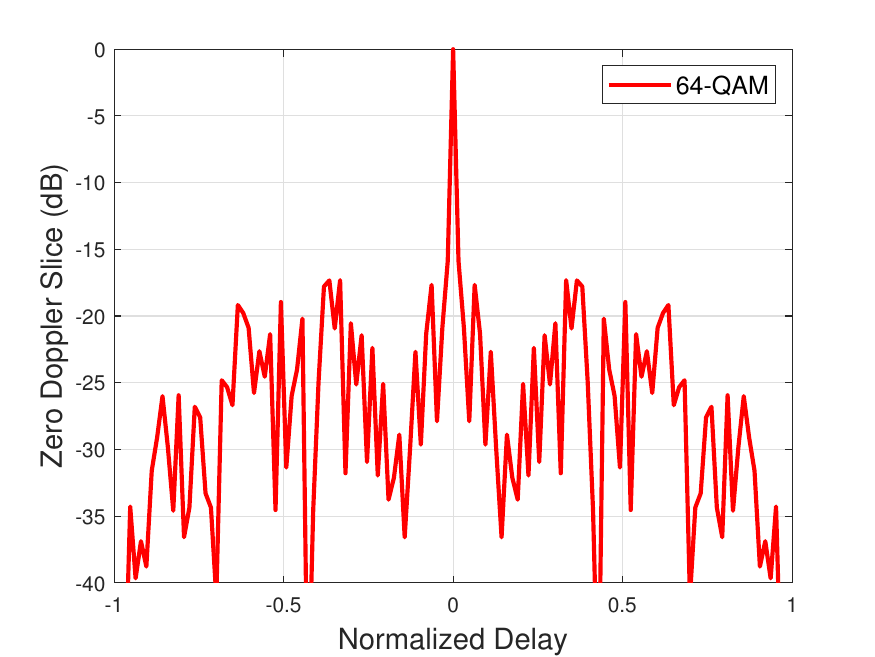}
		}
		\caption{AF of OFDM signals under uniformly distributed 64-QAM and 64-PSK modulations.}
		\label{fig1}
	\end{figure}

	Before presenting our PCS approaches with non-uniformly distributed constellations, we first evaluate the AF of OFDM communication signals for uniformly distributed 64-PSK and 64-QAM modulations. Notably, Fig. \ref{fig1a} unveils the presence of three peaks in the average AF of both 64-PSK and 64-QAM modulated signals, a phenomenon arising from the utilization of normalized axes with respect to $T_p$ and $B$, respectively. Therefore, two additional peaks appear on account of the Doppler ambiguity.
	As exhibited in Fig. \ref{fig1b}, it is evident that 64-PSK possesses significantly better average AF performance compared to 64-QAM, in terms of zero-Doppler slice (i.e., the autocorrelation function). The maximum gap of nearly 14 dB may be attributed to lower sidelobes of 64-PSK in zero-Doppler slice. Additionally, Fig. \ref{fig1c} and Fig. \ref{fig1d} depict the zero Doppler slices for a single realization of 64-PSK and 64-QAM modulated OFDM signal, respectively. The results demonstrate that both modulations result in random fluctuations in the AF, while 64-PSK is again superior to 64-QAM in terms of the lower random sidelobes.

	\begin{figure}[!t]
		\centering
		\includegraphics[width=2.80in]{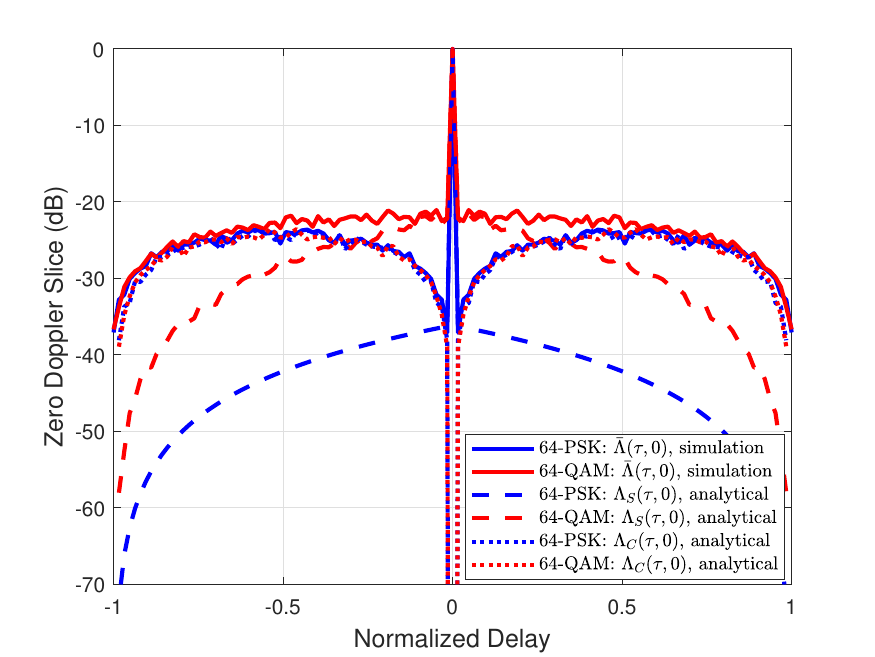}\\
		\caption{AF with simulated $\bar{\Lambda}(\tau,0)$, and the analytical $\bar{\Lambda}_\text{S}(\tau,0)$ and $\bar{\Lambda}_\text{C}(\tau,0)$.}
		\label{figx}
	\end{figure}

	\begin{figure}[!t]
		\centering
		\includegraphics[width=3.1in]{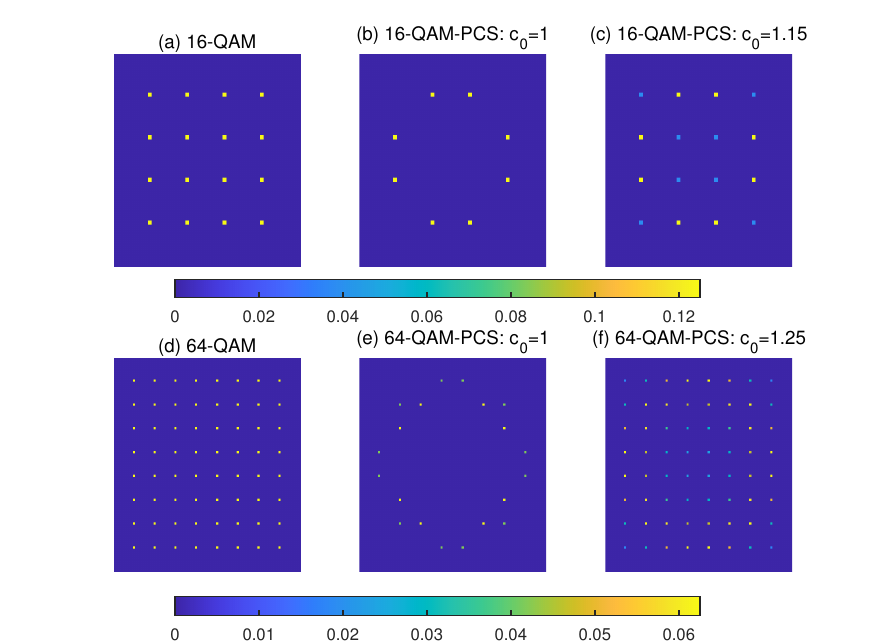}\\
		\caption{Heuristic PCS results with different $c_0$: 16-QAM and 64-QAM.}
		\label{fig2}
	\end{figure}

	\begin{figure}[!t]
		\centering
		\includegraphics[width=3.10in]{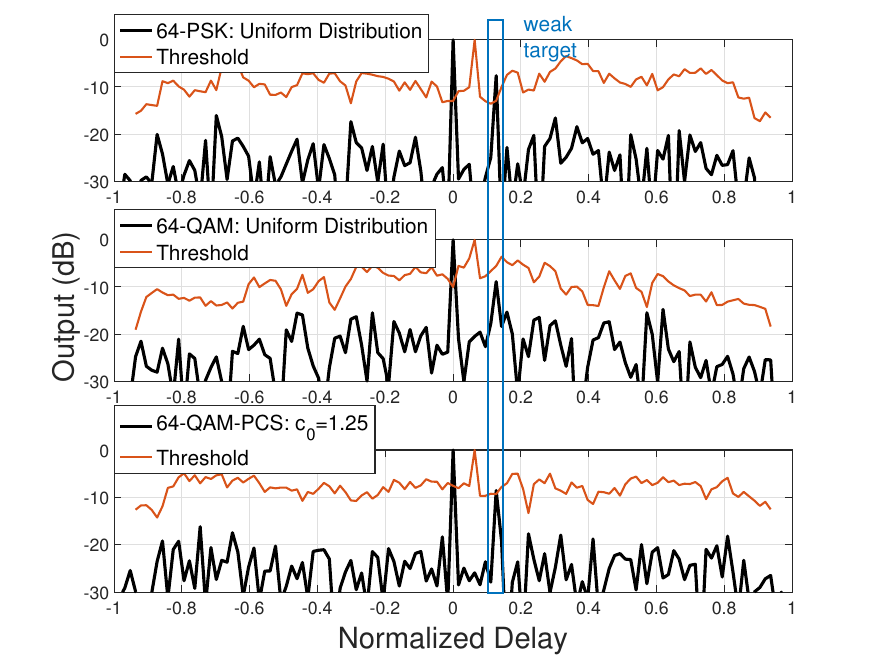}\\
		\caption{SO-CFAR results for single realizations of 64-QAM and 64-PSK.}
		\label{figxx}
	\end{figure}

	\begin{figure*}[!t]
		\subfigure[Probability of detection versus sensing SNR.] { \label{fig3}
			\includegraphics[width=0.626\columnwidth]{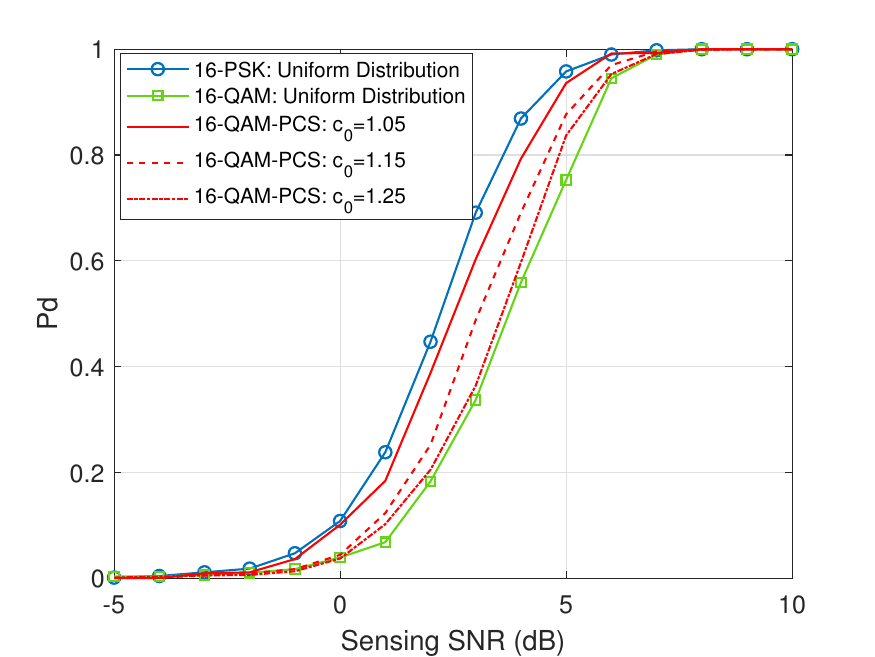}
		}
		\subfigure[16-QAM-PCS: AIR versus $c_0$.] { \label{fig4}
			\includegraphics[width=0.626\columnwidth]{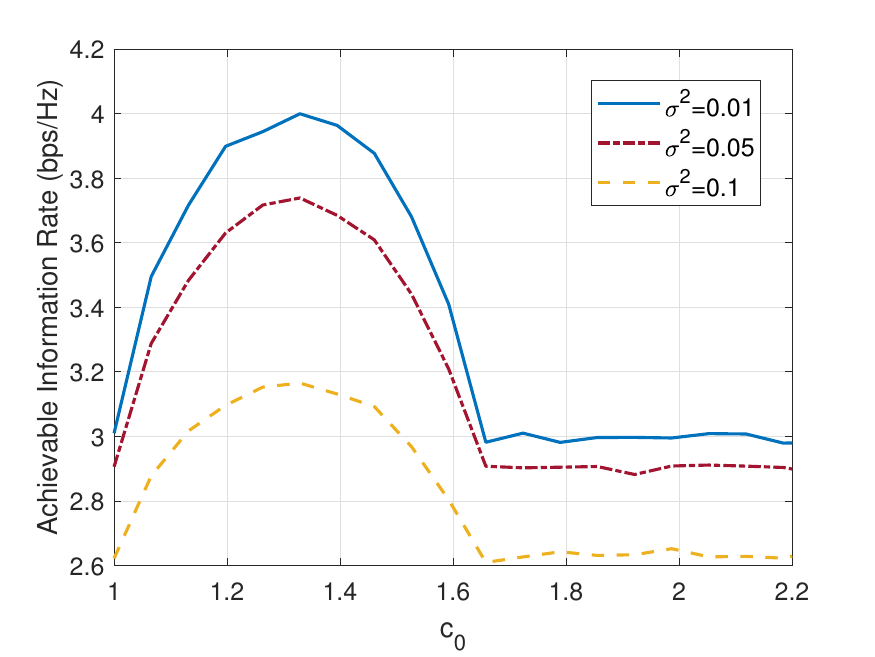}
		}
		\subfigure[AIR versus communication SNR.] { \label{fig6}
			\includegraphics[width=0.626\columnwidth]{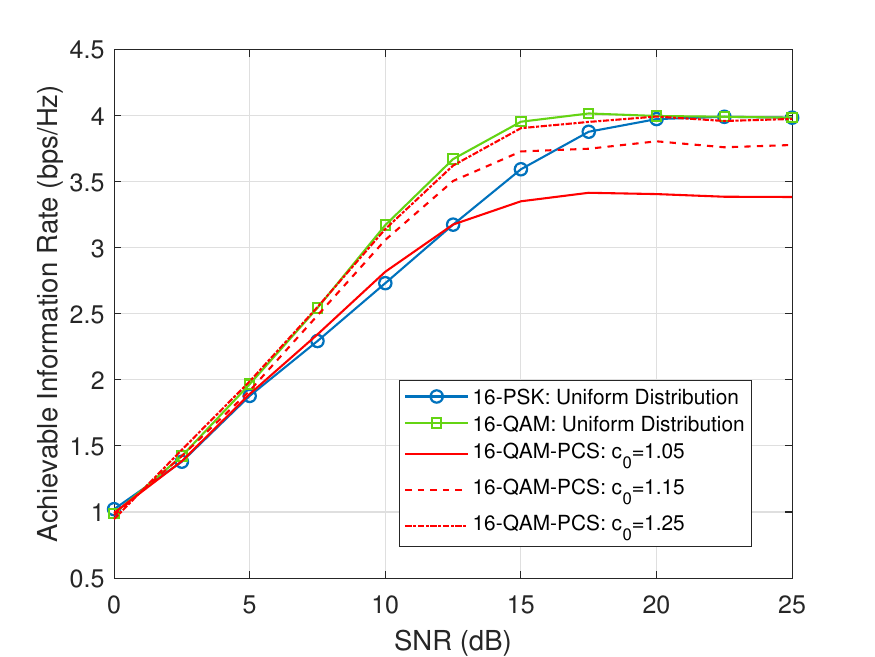}
		}
		\caption{16-QAM PCS: Heuristic PCS results for S\&C performance. The maximum AIR for 64-QAM is achieved at $c_0=1.32$.}
		\label{fig}
	\end{figure*}

	\begin{figure*}[!t]
		\subfigure[Probability of detection versus sensing SNR.] { \label{fig3x}
			\includegraphics[width=0.626\columnwidth]{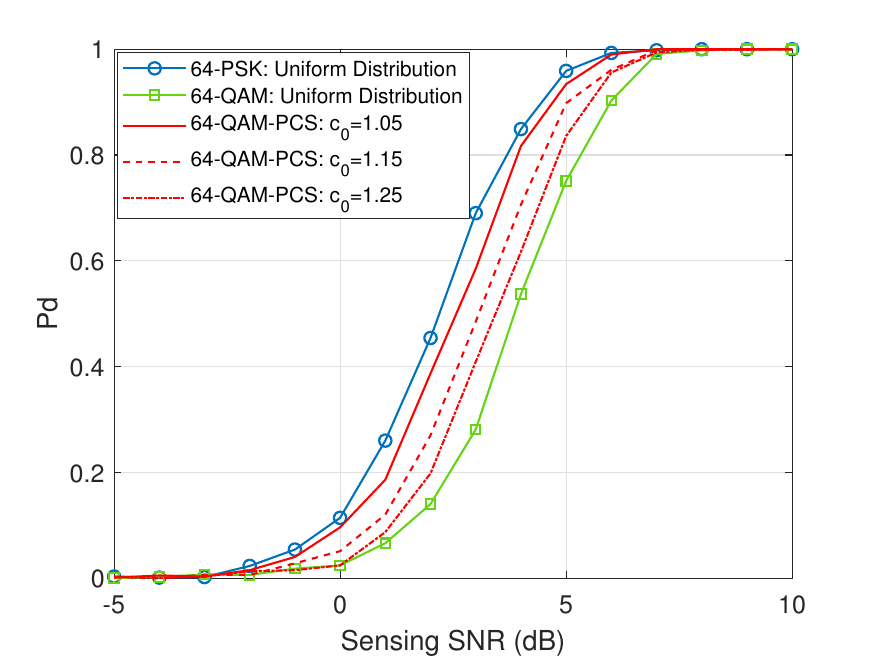}
		}
		\subfigure[64-QAM-PCS: AIR versus $c_0$.] { \label{fig4x}
			\includegraphics[width=0.626\columnwidth]{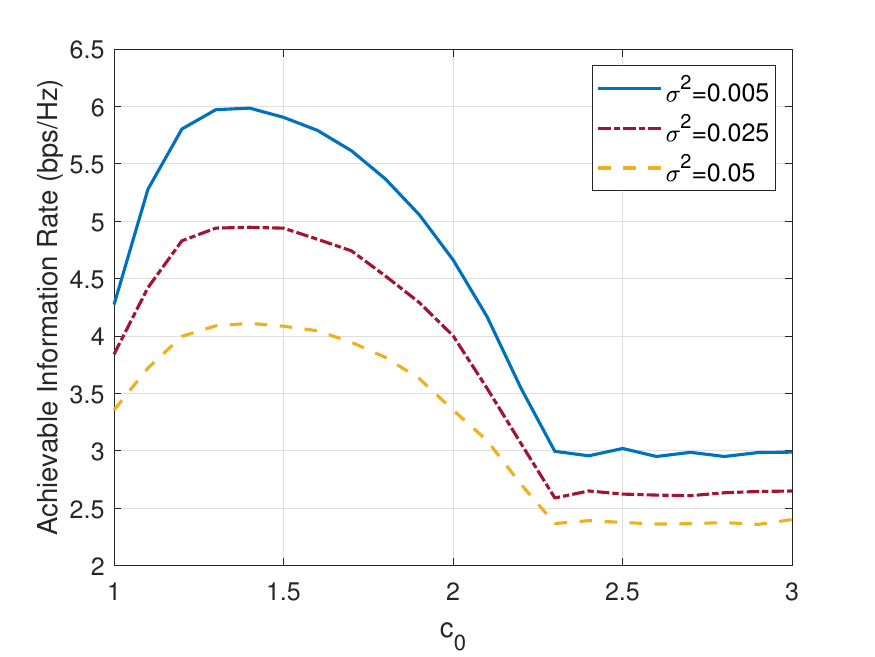}
		}
		\subfigure[AIR versus communication SNR.] { \label{fig6x}
			\includegraphics[width=0.626\columnwidth]{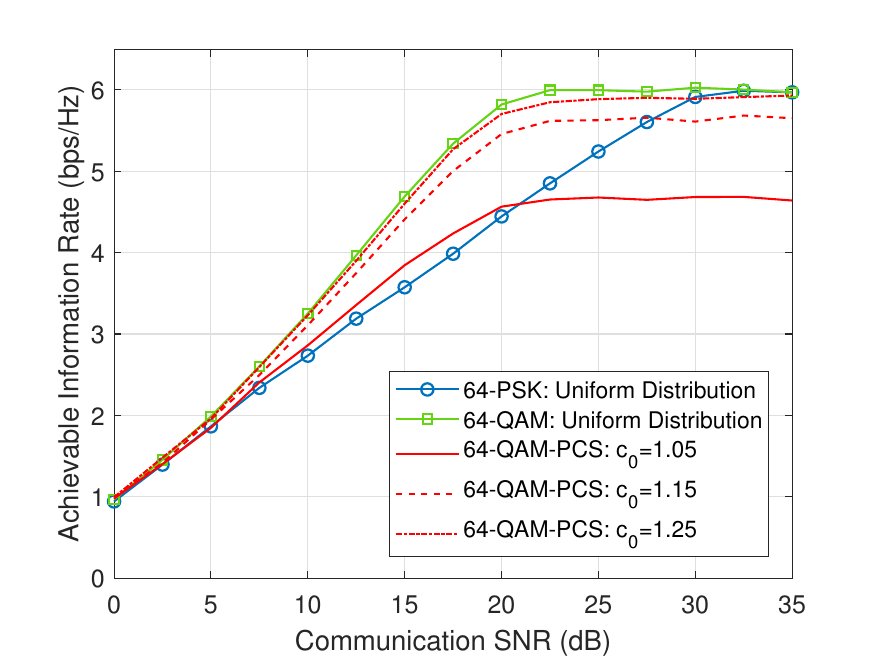}
		}
		\caption{64-QAM PCS: Heuristic PCS results for S\&C performance. The maximum AIR for 64-QAM is achieved at $c_0=1.3805$.}
		\label{fig7}
	\end{figure*}

	Next, in Fig. \ref{figx}, we present analytical results of $\bar{\Lambda}_\text{S}$ and $\bar{\Lambda}_\text{C}$, alongside the simulated autocorrelation functions of 64-QAM and 64-PSK, respectively. The sensing performance gap between 64-QAM and 64-PSK stems solely from $\bar{\Lambda}_\text{S}$. Consequently, the statistical characteristics of $\bar{\Lambda}_\text{C}$ bear no influence on the PCS method.


	\subsection{ISAC Performance with Heuristic PCS Approach}
	We next evaluate the effectiveness of the heuristic PCS approach in the context of QAM constellations (referred to as ``16/64-QAM-PCS''). As shown in Fig. \ref{fig2}(b) and (e), when $c_0=1$, the optimization model seeks for the solution of optimal sensing performance, in terms of the smallest fourth-moment of constellation amplitudes, i.e., $\mathbb{E}\left\{A^4_{x}\right\} - 1 =0$.
	For 16-QAM, PCS yields an output characterized by unit modulus points, constituting a pseudo 8-PSK. It is worth emphasizing that this constellation is not a real 8-PSK since the phase differences between adjacent points are not equal. On the other hand, in the case of 64-QAM, the PCS algorithm cannot find a constant modulus circle with unit power, thereby outputs two constant modulus circles proximate to the unit modulus circle, which corresponds to the fourth-moment value of $1.0363$. In addition, when $c_0$ increases, the PCS outputs the constellation with non-uniform input distribution, which balances between the best sensing ($c_0=1$) and the best communication (i.e., uniformly distributed QAM) performance.

	To assess the system-level sensing performance, and to illustrate the benefits of reducing sidelobes, we further consider a use case of detecting weak targets in the presence of strong self-interference (SI), which is applicable for practical ISAC scenarios operating in the in-band full duplex (IBFD) mode for short-range target detection \cite{barneto2019full}. To recover the weak target, the smallest of constant false alarm probability (SO-CFAR) detector \cite[Chapter 6.5.5]{richards2014fundamentals} is exploited, since the power of SI and noise elsewhere is not uniform. Thanks to such adaptive signal processing, the SI can be effectively excluded from the computation process of detection threshold. Throughout $5000$ Monte Carlo trials, the probability of false alarm ($P_\text{fa}$) is fixed as $10^{-4}$, and the weak target is allocated within the $8$th range cell, which is close to the SI at the $0$th range cell. The sensing signal-to-noise ratio (SNR) is defined as the power ratio between the weak target and the noise, while the power ratio between the SI and the noise is fixed as $10$ dB. Fig. \ref{figxx} illustrates the performance of SO-CFAR for random realizations of 64-QAM and 64-PSK, demonstrating that 64-PSK is more beneficial for weak target detection due to the lower random sidelobes of SI, while 64-QAM results in a higher threshold on the top of the weak target. Moreover, the detection performance of 64-QAM-PCS falls in between those of 64-QAM and 64-PSK. Nevertheless, the results of single realizations are not sufficiently convincing. To evaluate the average performance with $N_\text{MC}$ Monte Carlo trials, the probability of detection ($P_\text{d}$) versus the sensing SNR is portrayed in Fig. \ref{fig3} and Fig. \ref{fig3x}, which shows again that the proposed PCS approach achieves a performance tradeoff between 64-QAM and 64-PSK on an average sense. This indicates that the practical sensing performance (i.e., the $P_\text{d}$) may be flexibly adjusted by controlling the value of $c_0$.

	To evaluate the communication performance, we compute the AIR in an AWGN channel with Monte Carlo integrals in (\ref{Eq15}). In Fig. \ref{fig4} and Fig. \ref{fig4x}, the noise power $\sigma^2$ controls the receive communication SNR. For a high SNR case ($\sigma^2=0.01$), it is observed that the AIR reaches to the maximum values $4$bps/Hz in $c_0=1.32$ for 16-QAM-PCS and $6$ bps/Hz in $c_0=1.38$ for 64-QAM-PCS, corresponding to the entropy of the uniformly distributed 16-QAM and 64-QAM, namely,
	\begin{equation*}
		\begin{aligned}
			\sum\nolimits^{Q}_{q=1} A^4_q \cdot \mathbf{1}_Q/Q =
			\left\{
			\begin{array}{ll}
				1.32,   & \text{16-QAM}, \ $Q=16$, \\
				1.3805,  & \text{64-QAM}, \ $Q=64$. \\
			\end{array}
			\right.
		\end{aligned}
	\end{equation*}
	When $c_0=1$, the high-SNR AIR is 3bps/Hz, achieved by the pseudo 8-PSK constellation as shown in Fig. \ref{fig2}(b), indicating that the optimal sensing performance at this moment is attained at the price of 1bps/Hz rate loss. 
	Moreover, there is an obvious tradeoff between S\&C in the region of $c_0 \in [1,1.32]$, with known probability distributions on this curve, which is consistent with our analysis in Sec. IV-B. Notably, when $c_0>1.32$, the PCS is not uniform again, results in a declining AIR. In contrast, with $c_0>1.62$, the fourth moment of amplitudes reaches to its largest value and thus the AIR keeps constant as well.

	\begin{figure}[!t]
		\centering
		\includegraphics[width=3.0in]{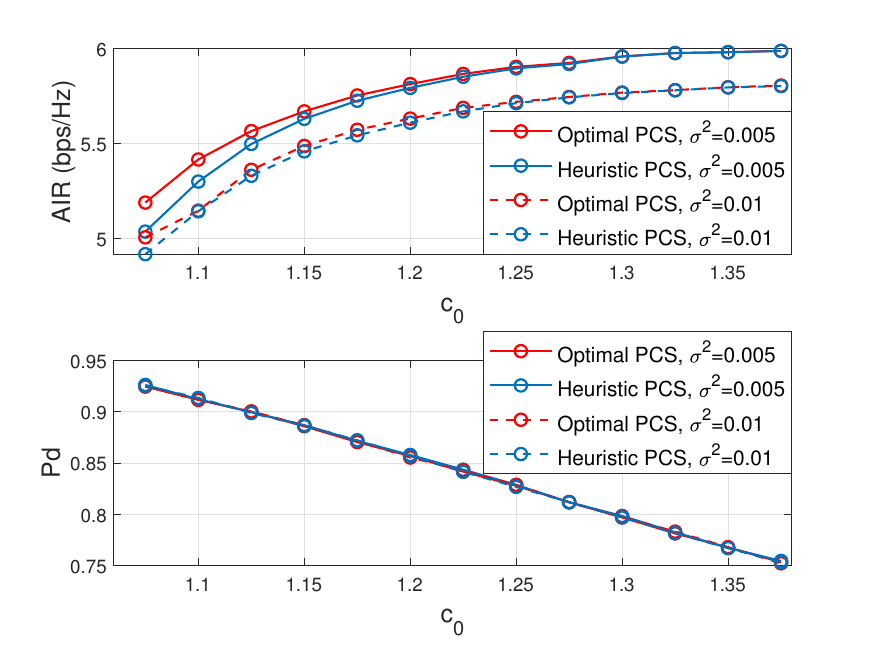}\\
		\caption{64-QAM PCS: AIR and $P_\text{d}$ versus $c_0$.}
		\label{figxxxx}
	\end{figure}

	\begin{figure}[!t]
		\centering
		\includegraphics[width=3.0in]{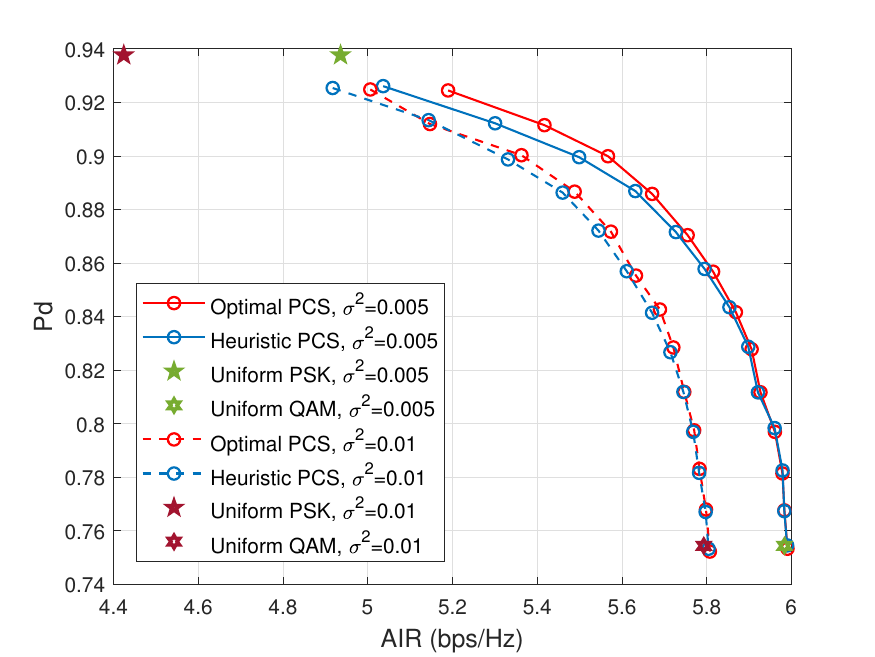}\\
		\caption{64-QAM PCS: S\&C tradeoff regions.}
		\label{figxxx}
	\end{figure}

	Fig. \ref{fig6} and Fig. \ref{fig6x} further reveal the advantages of the proposed PCS method across various SNR values. As anticipated in Fig. \ref{fig6}, both 16-QAM and 16-PSK reach to their capacity limit of 4bps/Hz, respectively, when the communication SNR is sufficiently high. However, in the relatively low SNR region, namely, for SNR = 10 dB in Fig. \ref{fig6}, a noticeable gap becomes apparent between 16-QAM and 16-PSK. Thanks to the heuristic PCS customization, one may achieve a transmission rate gain at the expense of sensing performance loss when comparing with its PSK counterparts, striking a scalable tradeoff between S\&C. Similar results for 64-QAM-PCS are also portrayed in Fig. \ref{fig6x}.

	\subsection{ISAC Performance with Optimal PCS Approach}
	
	We now evaluate the behavior of the proposed MBA algorithm for the optimal PCS optimization model, and compare its performance with the heuristic PCS approach. As demonstrated previously, the tradeoff regime between S\&C lies in:
	\begin{equation*}
		\begin{aligned}
			\left\{
			\begin{array}{ll}
				\text{16-QAM:} & c_0\in[1,1.32] \\
				\text{64-QAM:} & c_0\in[1.0363,1.3805] \\
			\end{array}
			\right.
		\end{aligned}
	\end{equation*}
	Therefore, we may only concentrate on these segments in the following. Notably, optimal PCS and heuristic PCS approaches have the same and unique solution for 16-QAM based constellation, which has been clarified in Sec. \ref{sec4}. That is why we consider only the 64-QAM case below.
	
	In Fig. \ref{figxxxx}, the two proposed PCS approaches are compared, in terms of their S\&C performance with varying $c_0$. On one hand, the optimal PCS approach slightly outperforms heuristic PCS counterpart in achieving higher AIR. This confirms that the heuristic PCS without involving communication metric may already approaches the optimal performance bound (i.e., the optimal PCS results) at a significantly reduced complexity. 
	On the other hand, these two approaches have almost identical $P_\text{d}$ due to the same fourth-moment of constellation amplitudes. We may conclude that optimal PCS achieves the Pareto frontier between S\&C performance, while the low-complexity heuristic PCS yields near-optimal solutions.
	
	Finally, the tradeoff regions for optimal PCS and heuristic PCS approaches are explicitly characterized in Fig. \ref{figxxx}, demonstrating that optimal PCS in general achieves the better performance over heuristic PCS, despite the relatively limited performance gain.
	The performance of conventional PSK and QAM with uniform constellation points are also depicted for comparison. Notably, the $P_\text{d}$ of PSK is larger over PCS results since PCS approaches for 64-QAM constellation cannot achieve a constant modulus shaping. However, the AIR of PSK is inferior to PCS enabled 64-QAM constellations, which coincides with the observation in Fig. \ref{fig6x}.
	Overall, the resulting S\&C performance boundaries confirm that both PCS approaches outperform the time-sharing strategy \cite{xiong2023fundamental}, corresponding to a line segment connected the sensing-optimal point to the communication-optimal point. Moreover, the optimal constellation probabilities on the performance boundaries are readily obtainable as well through the proposed PCS optimization approach. Different from the uniform distribution, both PCS approaches may scale the constellation probability distribution by controlling $c_0$, thereby affecting both S\&C performance simultaneously. 
	For higher-order QAM modulations or different noise levels, a look-up table of constellation probabilities may be conceived in an offline manner, indicating that the proposed ISAC signaling strategy may be configured in advance as per the practical service requirements.

	\section{Conclusion}\label{sec7}
	This article explored the tradeoff between S\&C using OFDM communication signals in monostatic ISAC systems. We first investigated the statistical characteristics of AF for OFDM signaling modulated by random symbols with arbitrary input distribution. Then, an optimal PCS approach was devised by maximizing the AIR, subject to the fourth-moment of constellation symbols, power, and probability constraints, which was numerically solved by a tailed MBA algorithm. Furthermore, a heuristic PCS approach was proposed by omitting the AIR metric, in an effort to actively control the fourth-moment of the constellation, and passively scale the communication performance by adjusting the input distribution. The main characteristics of our approaches lie in: 1) In contrast to the conventional uniform QAM modulation, the enhanced sensing performance owing to lower sidelobes of matched filtering, is attained at the expense of a certain AIR loss. 2) In contrast to the conventional uniform PSK modulation, the AIR in the low communication SNR region can be improved at the expense of certain sensing performance loss. Therefore, our approaches may strike a scalable performance tradeoff between S\&C. In addition, the heuristic PCS approach achieved very close performance to the S\&C tradeoff boundary achieved by the optimal PCS approach, while at a much lower computational complexity. Finally, both PCS approaches may be implemented offline in practical 6G ISAC applications.

	\appendices
	\section{Proof of Proposition 1}\label{appendix1}
	Substituting $s(t)$ into (\ref{Eq1}) yields
	\begin{equation}\label{Eq23}
		\begin{aligned}
			\Lambda(\tau,\nu) = & \sum\nolimits^{L-1}_{l_1=0}\sum\nolimits^{L-1}_{l_2=0} A_{l_1}A_{l_2}\exp\left(j\left(\psi_{l_1}-\psi_{l_2}\right)\right) \\ & \times \int^{\infty}_{-\infty} \phi_{l_1}(t)\phi^*_{l_2}(t-\tau) \exp\left(-j2\pi\nu t\right) dt.
		\end{aligned}
	\end{equation}
	The integral in (\ref{Eq23}) may be further recast as
	\begin{align}\label{Eq24}
		\int^{\infty}_{-\infty} & \phi_{l_1}(t) \phi^*_{l_2}(t-\tau) \exp\left(-j2\pi\nu t\right) dt 
		= \exp\left(j2\pi l_2\Delta f \tau\right) \nonumber
		\\ & \times \int^{T_\text{max}}_{T_\text{min}}  \exp\left\{j2\pi \left[\left(l_1-l_2\right)\Delta f -\nu\right] t\right\} dt.
	\end{align}

	To proceed, we rely on the following equation:
	\begin{equation*}
		\begin{aligned}
			\int^{T_\text{max}}_{T_\text{min}} \exp\left(j2\pi ft\right)  dt
			= & T_\text{diff}\text{sinc}\left(f T_\text{diff}\right) \exp\left(j2\pi f T_\text{avg}\right).
		\end{aligned}
	\end{equation*}
	Then it is straightforward to reformulate (\ref{Eq24}) as
	\begin{align}\label{Eq25}
		& \int^{\infty}_{-\infty} \phi_{l_1}(t)\phi^*_{l_2}(t-\tau) \exp\left(-j2\pi\nu t\right)  dt \nonumber
		\\ & = T_\text{diff} \text{sinc}\left\{ \left[\left(l_1-l_2\right)\Delta f -\nu\right] T_\text{diff} \right\} \\ & \quad \times  \exp\left\{j2\pi \left[\left(\left(l_1-l_2\right)\Delta f -\nu\right) T_\text{avg} + l_2\Delta f \tau \right]  \right\} . \nonumber
	\end{align}
	Inserting (\ref{Eq25}) into (\ref{Eq23}) proves Proposition 1.

	\section{Proof of Proposition 2}\label{appendix2}
	The derivation of $\mathbb{E}\left(\Lambda_\text{S}\Lambda^*_\text{C}\right)$ is shown as:
	\begin{equation}
	\begin{aligned}\label{lambdasc}
		\mathbb{E}\left\{\Lambda_\text{S} \Lambda^*_\text{C}  \right\} = &  {K_0}  \sum\nolimits^{L-1}_{l=0}  \sum\nolimits^{L-1}_{l_1=0} \sum\nolimits^{L-1}_{l_2=0 \atop l_2\neq l_1}  \mathbb{E}\{ A^2_{l} A_{l_1}A_{l_2} \\ & \times \exp(j(\psi_{l_1}-\psi_{l_2})) \} {K_{l,l_1,l_2}},
	\end{aligned}
	\end{equation}	
	where $K_0$ and $K_{l,l_1,l_2}$ are constant with respect to the input distribution of constellation. 

	Then we have $\mathbb{E}\left(\Lambda_\text{S}\Lambda^*_\text{C}\right) = 0$,
	due to the fact that
	\begin{equation*}
		\begin{aligned}	
			& \mathbb{E}\left\{ A^2_{l}A_{l_1}A_{l_2}\exp\left(-j\left(\psi_{l_1}-\psi_{l_2}\right)\right)\right\}
			=  \\ & \left\{
			\begin{array}{ll}
				{\mathbb{E}\left\{A^3_{l_1}e^{-j\psi_{l_1}}\right\} \mathbb{E}\left\{A_{l_2}e^{j\psi_{l_2}}\right\}}=0,  & l=l_1\neq l_2 \\
				{\mathbb{E}\left\{A^3_{l_2}e^{j\psi_{l_2}}\right\} \mathbb{E}\left\{A_{l_1}e^{-j\psi_{l_1}}\right\}}=0, & l=l_2\neq l_1 \\
				{\mathbb{E}\left\{A^2_{l}\right\} \mathbb{E}\left\{A_{l_1}e^{j\psi_{l_1}}\right\}
				\mathbb{E}\left\{A_{l_2}e^{-j\psi_{l_2}}\right\}}=0, & l\neq l_1\neq l_2 
			\end{array}
			\right.
		\end{aligned}
	\end{equation*}
	where the derivation exploits the rule that $\mathbb{E}\{ABC\}=\mathbb{E}\{A\}\mathbb{E}\{B\}\mathbb{E}\{C\}$, if $A$, $B$ and $C$ are independent. For example, when $l_1\neq l_2$ we have $\mathbb{E}\left\{A_{l_1}A_{l_2}e^{-j(\psi_{l_1}-\psi_{l_2})}\right\} = {\mathbb{E}\left\{A_{l_1}e^{-j\psi_{l_1}}\right\}\mathbb{E}\left\{A_{l_2}e^{j\psi_{l_2}}\right\}}$ since $A_{l_1}e^{-j\psi_{l_1}}$ and $A_{l_2}e^{j\psi_{l_2}}$ are generated from i.i.d. input symbols in the constellation. Moreover, all three cases equal to zero due to that $\mathbb{E}\left\{A_{l}  \exp\left(j\psi_{l}\right) \right\} = 0$, which has been shown in Assumption 1. Benefiting from this, it is straightforward to see that $\mathbb{E}\{\Lambda_\text{C}(\tau,\nu)\}=0$. Proposition 2 is thus proved.

	\section
	{Proof of Proposition 3}\label{appendix3}
	Following a similar derivation procedure of $\sigma^2_\text{S}$, we formulate the variance of $\tilde{\Lambda}_\text{S}$ as
		\begin{align}
			& \tilde{\sigma}^2_\text{S} = \mathbb{E}[|\tilde{\Lambda}_\text{S}|^2] - |\mathbb{E}[\tilde{\Lambda}_\text{S}]|^2  \nonumber
			\\ & = T^2_\text{diff} \text{sinc}^2\left( - \nu T_\text{diff} \right) \sum^{N-1}_{n=0}\sum^{L-1}_{l=0} \sum^{N-1}_{n'=0}\sum^{L-1}_{l'=0} \left\{\mathbb{E}\left\{A^2_{n,l}A^2_{n',l'}\right\} - 1\right\}  \nonumber
			\\ & \quad \times \exp\left\{j2\pi \left[\left(l-l'\right)\Delta f \tau - \left(n-n'\right)\nu T_p \right]\right\} 
		\end{align} 
	In accordance with
	\begin{equation*}
		\begin{aligned}
			&  \!\!\!\!\! \mathbb{E}\left\{A^2_{n,l} A^2_{n',l'}\right\}= \left\{
			\begin{array}{lc}
				{ \!\!\! \mathbb{E}\left\{A^4_{n,l}\right\} }, & \!\!\!\!\!\! \!\!\!\!\!\! n=n',l=l' \\
				\!\!\! \mathbb{E}\left\{A^2_{n,l}\right\} \mathbb{E}\left\{A^2_{n',l'}\right\}=1,  & \!\!\!\!  \text{Otherwise}
			\end{array}
			\right.
		\end{aligned}
	\end{equation*}
	we may further simplify $\tilde{\sigma}^2_\text{S}\left(\tau,\nu\right)$ as
	\begin{equation*}
		\begin{aligned}
			\tilde{\sigma}^2_\text{S}  = T^2_\text{diff} \text{sinc}^2\left( - \nu T_\text{diff} \right) \sum\nolimits^{N-1}_{n=0} \sum\nolimits^{L-1}_{l=0} \left({\mathbb{E}\left\{A^4_{n,l}\right\} - 1} \right).
		\end{aligned} 
	\end{equation*}
	Compared with (\ref{Eq6}), it is evident that $\tilde{\sigma}^2_\text{S} \left(\tau,\nu\right)$ is the variance accumulation of $N$ symbols. Therefore, Proposition 3 is proved.
	Just for discussion, if the constellation varies with the symbol and the subcarrier, namely, the subscripts $n$ and $l$, then more DoFs can be exploited to scale the signaling randomness. This can be left as our future work.
	
	\section
	{Proof of Proposition 4}\label{appendix4}
	Likewise, we derive the variance of $\tilde{\Lambda}_\text{C}$ as
	\begin{equation}
		\begin{aligned}
			\tilde{\sigma}^2_\text{C} =  \mathbb{E}\{|\tilde{\Lambda}_\text{C}|^2\} - |\mathbb{E}\{\tilde{\Lambda}_\text{C}\}|^2.
		\end{aligned}
	\end{equation}
	Firstly,  $\mathbb{E}\{\tilde{\Lambda}_\text{C}\} = 0$ can be readily verified. We therefore only need to derive  $\mathbb{E}\{|\tilde{\Lambda}_\text{C}|^2\}$.
	For brevity, we denote the two terms in (\ref{Eq13}) by $Z_1$ and $Z_2$, and subsequently reformulate (\ref{Eq13}) as
	\begin{equation}
		\begin{aligned}
			\tilde{\sigma}^2_\text{C} = \mathbb{E}\{|Z_1|^2\}+\mathbb{E}\{|Z_2|^2\} 
			-2\Re\{\mathbb{E}\{Z_1Z^*_2\}\},
		\end{aligned}
	\end{equation}
	where 
	\begin{align}
		\mathbb{E} \left\{\left|Z_1\right|^2\right\} = & \sum^{N-1}_{n=0} \sum^{L-1}_{l_1=0} \sum^{L-1}_{l_2=0 \atop l_2\neq l_1} \sum^{N-1}_{n'=0} \sum^{L-1}_{l'_1=0} \sum^{L-1}_{l'_2=0 \atop l'_2\neq l'_1} 
		\mathbb{E}\big\{ A_{n,l_1} A_{n,l_2}  \nonumber \\ & \times  A_{n',l'_1}  A_{n',l'_2} e^{j(\psi_{n,l_1}-\psi_{n,l_2}-\psi_{n',l'_1} +\psi_{n',l'_2})} \big\} \nonumber \\ & \times R_1\left(n,l_1,l_2\right)R^*_1\left(n',l'_1,l'_2\right),
		\\
		\mathbb{E} \left\{\left|Z_2\right|^2\right\} = & \sum^{N-1}_{n_1=0} \sum^{N-1}_{n_2=0 \atop n_2 \neq n_1} \sum^{L-1}_{l_1=0} \sum^{L-1}_{l_2=0} \sum^{N-1}_{n'_1=0} \sum^{N-1}_{n'_2=0 \atop n'_2 \neq n'_1} \sum^{L-1}_{l'_1=0} \sum^{L-1}_{l'_2=0} \mathbb{E}\big\{  \nonumber \\ &  \!\!\!\!\!\!\!\!\!\! \!\!\!\!\!\!\!\!\!\!  A_{n_1,l_1} A_{n_2,l_2}  A_{n'_1,l'_1} A_{n'_2,l'_2}  e^{j(\psi_{n_1,l_1}-\psi_{n_2,l_2} -\psi_{n'_1,l'_1} +\psi_{n'_2,l'_2})} \big\} \nonumber \\ &  \times R_2\left(n_1,n_2,l_1,l_2\right) R^*_2\left(n'_1,n'_2,l'_1,l'_2\right), 			
		\\
		\mathbb{E} \left\{Z_1Z^*_2\right\} = & 
		\sum^{N-1}_{n=0} \sum^{L-1}_{l_1=0} \sum^{L-1}_{l_2=0 \atop l_2\neq l_1} 
		\sum^{N-1}_{n'_1=0} \sum^{N-1}_{n'_2=0 \atop n'_2 \neq n'_1} \sum^{L-1}_{l'_1=0} \sum^{L-1}_{l'_2=0}
		\mathbb{E}\big\{
		A_{n,l_1} A_{n,l_2} \nonumber \\ &  \times  A_{n'_1,l'_1}   A_{n'_2,l'_2} e^{j(\psi_{n,l_1}-\psi_{n,l_2} - \psi_{n'_1,l'_1} + \psi_{n'_2,l'_2})}
		\big\} \nonumber \\ & \times R_1\left(n,l_1,l_2\right)R^*_2\left(n'_1,n'_2,l'_1,l'_2\right). 
	\end{align}
	For further simplifications, we note the following facts that 
	\begin{flalign*}	
		\quad & \mathbb{E}\big\{A_{n,l_1}A_{n,l_2}A_{n',l'_1}A_{n',l'_2} \nonumber \\ & \quad \times \exp\left(j\left(\psi_{n,l_1}-\psi_{n,l_2}-\psi_{n',l'_1}+\psi_{n',l'_2}\right)\right) \big\}  \nonumber
		= \\ & \left\{
		\begin{array}{ll}
			\mathbb{E}\left\{ A^2_{n,l_1}A^2_{n',l'_1} \right\}, &   l_1=l_2,l'_1=l'_2 \\
			\mathbb{E}\left\{ A^2_{n,l_1}A^2_{n,l_2} \right\},   &   l_1=l'_1,l_2=l'_2,n=n' \\
			0, &    \text{otherwise}
		\end{array}
		\right. &
	\end{flalign*}
	\begin{flalign*}	
		\quad & \mathbb{E}\big\{A_{n_1,l_1}A_{n_2,l_2}A_{n'_1,l'_1}A_{n'_2,l'_2} \nonumber \\ & \quad  \times \exp\left(j\left(\psi_{n_1,l_1}-\psi_{n_2,l_2}-\psi_{n'_1,l'_1}+\psi_{n'_2,l'_2}\right)\right) \big\}  \nonumber
		=  \\ & \left\{
		\begin{array}{ll}
			\mathbb{E}\left\{ A^2_{n_1,l_1}A^2_{n'_1,l'_1} \right\}, &  l_1=l_2,l'_1=l'_2,n_1=n_2,n'_1=n'_2 \\
			\mathbb{E}\left\{ A^2_{n_1,l_1}A^2_{n_2,l_2} \right\}, &   l_1=l'_1,l_2=l'_2,n_1=n'_1,n_2=n'_2 \\
			0,  &   \text{otherwise}
		\end{array}
		\right.&
	\end{flalign*}
	\begin{flalign*}	
		\quad & \mathbb{E}\big\{A_{n,l_1}A_{n,l_2}A_{n'_1,l'_1}A_{n'_2,l'_2} \nonumber \\ & \quad  \times \exp(j(\psi_{n,l_1}-\psi_{n,l_2}-\psi_{n'_1,l'_1}+\psi_{n'_2,l'_2})) \big\}  \nonumber
		= \\ & \left\{
		\begin{array}{ll}
			\mathbb{E}\left\{ A^2_{n,l_1}A^2_{n',l'_1} \right\},  &    l_1=l_2,l'_1=l'_2 \\
			\mathbb{E}\left\{ A^2_{n,l_1}A^2_{n,l_2} \right\},  &   l_1=l'_1,l_2=l'_2,n=n'_1=n'_2 \\
			0,  &    \text{otherwise}
		\end{array}
		\right.&
	\end{flalign*}
	Therefore, the expectations can be significantly simplified as
	\begin{align*}
		\mathbb{E}\{\left|Z_1\right|^2\} & = \sum\nolimits^{N-1}_{n=0} \sum\nolimits^{L-1}_{l_1=0} \sum\nolimits^{L-1}_{l_2=0 \atop l_2\neq l_1} 
		|R_1(n,l_1,l_2)|^2, \\
		\mathbb{E}\{\left|Z_2\right|^2\} & = \sum^{N-1}_{n_1=0} \sum^{N-1}_{n_2=0 \atop n_2 	\neq n_1} \sum^{L-1}_{l_1=0} \sum^{L-1}_{l_2=0}  |R_2(n_1,n_2,l_1,l_2) |^2,			
		\\
		\mathbb{E}\{Z_1Z^*_2\} & = 0.
	\end{align*}
	Finally, one can obtain
	\begin{equation*}
		\begin{aligned}
			\tilde{\sigma}^2_\text{C} & = \sum\nolimits^{N-1}_{n=0} \sum\nolimits^{L-1}_{l_1=0} \sum\nolimits^{L-1}_{l_2=0 \atop l_2\neq l_1} 
			|R_1(n,l_1,l_2)|^2 
			\\ & +
			\sum\nolimits^{N-1}_{n_1=0} \sum\nolimits^{N-1}_{n_2=0 \atop n_2 \neq n_1} \sum\nolimits^{L-1}_{l_1=0} \sum\nolimits^{L-1}_{l_2=0}  |R_2(n_1,n_2,l_1,l_2) |^2. 
		\end{aligned}
	\end{equation*}
	Similarly, this result suggests that $\tilde{\sigma}^2_\text{C}$ is a fixed constant for different constellations. 
	Therefore, Proposition 4 is proved.

	\ifCLASSOPTIONcaptionsoff
	\newpage
	\fi
	
	\bibliographystyle{IEEEtran}
	\bibliography{reference}

\begin{thebibliography}{10}
\providecommand{\url}[1]{#1}
\csname url@samestyle\endcsname
\providecommand{\newblock}{\relax}
\providecommand{\bibinfo}[2]{#2}
\providecommand{\BIBentrySTDinterwordspacing}{\spaceskip=0pt\relax}
\providecommand{\BIBentryALTinterwordstretchfactor}{4}
\providecommand{\BIBentryALTinterwordspacing}{\spaceskip=\fontdimen2\font plus
\BIBentryALTinterwordstretchfactor\fontdimen3\font minus
  \fontdimen4\font\relax}
\providecommand{\BIBforeignlanguage}[2]{{%
\expandafter\ifx\csname l@#1\endcsname\relax
\typeout{** WARNING: IEEEtran.bst: No hyphenation pattern has been}%
\typeout{** loaded for the language `#1'. Using the pattern for}%
\typeout{** the default language instead.}%
\else
\language=\csname l@#1\endcsname
\fi
#2}}
\providecommand{\BIBdecl}{\relax}
\BIBdecl

\bibitem{PCSglobecom}
Z.~Du, F.~Liu, Y.~Xiong, T.~X. Han, W.~Yuan, Y.~Cui, C.~Yao, and Y.~C. Eldar,
  ``Probabilistic constellation shaping for {OFDM}-based {ISAC} signaling,'' in
  \emph{Proc. IEEE Global Commun. Conf (GLOBECOM)}.\hskip 1em plus 0.5em minus
  0.4em\relax IEEE, 2023. Available: http://arxiv.org/abs/2310.18090.

\bibitem{liu2023integrated}
F.~Liu, C.~Masouros, and Y.~C. Eldar, \emph{Integrated Sensing and
  Communications}.\hskip 1em plus 0.5em minus 0.4em\relax Springer Nature,
  2023.

\bibitem{cui2021integrating}
Y.~Cui, F.~Liu, X.~Jing, and J.~Mu, ``Integrating sensing and communications
  for ubiquitous {I}o{T}: Applications, trends, and challenges,'' \emph{IEEE
  Netw}, vol.~35, no.~5, pp. 158--167, 2021.

\bibitem{liu2022integrated}
F.~Liu, Y.~Cui, C.~Masouros, J.~Xu, T.~X. Han, Y.~C. Eldar, and S.~Buzzi,
  ``Integrated sensing and communications: Toward dual-functional wireless
  networks for 6{G} and beyond,'' \emph{IEEE J. Sel Areas Commun}, vol.~40,
  no.~6, pp. 1728--1767, 2022.

\bibitem{ITU-R}
ITU-R, ``Framework and overall objectives of the future development of {IMT}
  for 2030 and beyond,'' Draft new recommendation, June, 2023.

\bibitem{du2023towards}
Z.~Du, F.~Liu, Y.~Li, W.~Yuan, Y.~Cui, Z.~Zhang, C.~Masouros, and B.~Ai,
  ``Towards {ISAC}-empowered vehicular networks: Framework, advances, and
  opportunities,'' \emph{arXiv preprint arXiv:2305.00681}, 2023.

\bibitem{dong2023communication}
F.~Dong, F.~Liu, S.~Lu, Y.~Xiong, Q.~Zhang, and Z.~Feng,
  ``Communication-assisted sensing in 6{G} networks,'' \emph{arXiv preprint
  arXiv:2311.07157}, 2023.

\bibitem{saddik2007ultra}
G.~N. Saddik, R.~S. Singh, and E.~R. Brown, ``Ultra-wideband multifunctional
  communications/radar system,'' \emph{IEEE Trans. Microw Theory. Techn},
  vol.~55, no.~7, pp. 1431--1437, 2007.

\bibitem{hassanien2015dual}
A.~Hassanien, M.~G. Amin, Y.~D. Zhang, and F.~Ahmad, ``Dual-function
  radar-communications: Information embedding using sidelobe control and
  waveform diversity,'' \emph{IEEE Trans. Signal Process}, vol.~64, no.~8, pp.
  2168--2181, 2015.

\bibitem{huang2020majorcom}
T.~Huang, N.~Shlezinger, X.~Xu, Y.~Liu, and Y.~C. Eldar, ``{MAJ}o{RC}om: A
  dual-function radar communication system using index modulation,'' \emph{IEEE
  Trans. Signal Process}, vol.~68, pp. 3423--3438, 2020.

\bibitem{sturm2011waveform}
C.~Sturm and W.~Wiesbeck, ``Waveform design and signal processing aspects for
  fusion of wireless communications and radar sensing,'' \emph{Proc. IEEE},
  vol.~99, no.~7, pp. 1236--1259, 2011.

\bibitem{kumari2017ieee}
P.~Kumari, J.~Choi, N.~Gonz{\'a}lez-Prelcic, and R.~W. Heath, ``{IEEE} 802.11
  ad-based radar: An approach to joint vehicular communication-radar system,''
  \emph{IEEE Trans. Veh Technol}, vol.~67, no.~4, pp. 3012--3027, 2017.

\bibitem{liu2018mu}
F.~Liu, C.~Masouros, A.~Li, H.~Sun, and L.~Hanzo, ``{MU-MIMO} communications
  with {MIMO} radar: From co-existence to joint transmission,'' \emph{IEEE
  Trans. Wireless Commun}, vol.~17, no.~4, pp. 2755--2770, 2018.

\bibitem{sen2010adaptive}
S.~Sen and A.~Nehorai, ``Adaptive {OFDM} radar for target detection in
  multipath scenarios,'' \emph{IEEE Trans. Signal Process}, vol.~59, no.~1, pp.
  78--90, 2010.

\bibitem{sen2012ofdm}
S.~Sen, ``{OFDM} radar space-time adaptive processing by exploiting
  spatio-temporal sparsity,'' \emph{IEEE Trans. Signal Process}, vol.~61,
  no.~1, pp. 118--130, 2012.

\bibitem{sen2010ofdm}
S.~Sen and A.~Nehorai, ``{OFDM MIMO} radar with mutual-information waveform
  design for low-grazing angle tracking,'' \emph{IEEE Trans. Signal Process},
  vol.~58, no.~6, pp. 3152--3162, 2010.

\bibitem{bicua2016generalized}
M.~Bic{\u{a}} and V.~Koivunen, ``Generalized multicarrier radar: Models and
  performance,'' \emph{IEEE Trans. Signal Process}, vol.~64, no.~17, pp.
  4389--4402, 2016.

\bibitem{shi2017power}
C.~Shi, F.~Wang, M.~Sellathurai, J.~Zhou, and S.~Salous, ``Power
  minimization-based robust {OFDM} radar waveform design for radar and
  communication systems in coexistence,'' \emph{IEEE Trans. Signal Process},
  vol.~66, no.~5, pp. 1316--1330, 2017.

\bibitem{bicua2018radar}
M.~Bic{\u{a}} and V.~Koivunen, ``Radar waveform optimization for target
  parameter estimation in cooperative radar-communications systems,''
  \emph{IEEE Trans. Aerosp. Electron. Syst}, vol.~55, no.~5, pp. 2314--2326,
  2018.

\bibitem{du2020distributed}
Z.~Du, Z.~Zhang, and W.~Yu, ``Distributed target detection in communication
  interference and noise using {OFDM} radar,'' \emph{IEEE Commun. Lett},
  vol.~25, no.~2, pp. 598--602, 2020.

\bibitem{xiong2023fundamental}
Y.~Xiong, F.~Liu, Y.~Cui, W.~Yuan, T.~X. Han, and G.~Caire, ``On the
  fundamental tradeoff of integrated sensing and communications under
  {G}aussian channels,'' \emph{IEEE Trans. Inf. Theory}, 2023.

\bibitem{berger2010signal}
C.~R. Berger, B.~Demissie, J.~Heckenbach, P.~Willett, and S.~Zhou, ``Signal
  processing for passive radar using {OFDM} waveforms,'' \emph{IEEE J. Sel
  Topics Signal Process}, vol.~4, no.~1, pp. 226--238, 2010.

\bibitem{hakobyan2017novel}
G.~Hakobyan and B.~Yang, ``A novel intercarrier-interference free signal
  processing scheme for {OFDM} radar,'' \emph{IEEE Trans. Veh Technol},
  vol.~67, no.~6, pp. 5158--5167, 2017.

\bibitem{zhang2020joint}
F.~Zhang, Z.~Zhang, W.~Yu, and T.-K. Truong, ``Joint range and velocity
  estimation with intrapulse and intersubcarrier doppler effects for
  {OFDM}-based {R}ad{C}om systems,'' \emph{IEEE Trans. Signal Process},
  vol.~68, pp. 662--675, 2020.

\bibitem{keskin2021mimo}
M.~F. Keskin, H.~Wymeersch, and V.~Koivunen, ``{MIMO-OFDM} joint
  radar-communications: Is {ICI} friend or foe?'' \emph{IEEE J. Sel Topics
  Signal Process}, vol.~15, no.~6, pp. 1393--1408, 2021.

\bibitem{bocherer2014probabilistic}
G.~B{\"o}cherer, ``Probabilistic signal shaping for bit-metric decoding,'' in
  \emph{Proc. IEEE Int. Symp. Inf. Theory (ISIT)}.\hskip 1em plus 0.5em minus
  0.4em\relax IEEE, 2014, pp. 431--435.

\bibitem{bocherer2015bandwidth}
G.~B{\"o}cherer, F.~Steiner, and P.~Schulte, ``Bandwidth efficient and
  rate-matched low-density parity-check coded modulation,'' \emph{IEEE Trans.
  Commun}, vol.~63, no.~12, pp. 4651--4665, 2015.

\bibitem{cho2019probabilistic}
J.~Cho and P.~J. Winzer, ``Probabilistic constellation shaping for optical
  fiber communications,'' \emph{J. Lightw. Technol}, vol.~37, no.~6, pp.
  1590--1607, 2019.

\bibitem{steiner2020coding}
F.~Steiner, ``Coding for higher-order modulation and probabilistic shaping,''
  Ph.D. dissertation, Technische Universit{\"a}t M{\"u}nchen, 2020.

\bibitem{schulte2015constant}
P.~Schulte and G.~B{\"o}cherer, ``Constant composition distribution matching,''
  \emph{IEEE Trans. Inf. Theory}, vol.~62, no.~1, pp. 430--434, 2015.

\bibitem{tigrek2012ofdm}
R.~F. Tigrek, W.~J. De~Heij, and P.~Van~Genderen, ``{OFDM} signals as the radar
  waveform to solve {D}oppler ambiguity,'' \emph{IEEE Trans. Aerosp. Electron.
  Syst}, vol.~48, no.~1, pp. 130--143, 2012.

\bibitem{woodward2014probability}
P.~M. Woodward, \emph{Probability and information theory, with applications to
  radar: international series of monographs on electronics and
  instrumentation}.\hskip 1em plus 0.5em minus 0.4em\relax Elsevier, 2014,
  vol.~3.

\bibitem{sen2009adaptive}
S.~Sen and A.~Nehorai, ``Adaptive design of {OFDM} radar signal with improved
  wideband ambiguity function,'' \emph{IEEE Trans. Signal Process}, vol.~58,
  no.~2, pp. 928--933, 2009.

\bibitem{carmon2015comparison}
Y.~Carmon, S.~Shamai, and T.~Weissman, ``Comparison of the achievable rates in
  {OFDM} and single carrier modulation with {IID} inputs,'' \emph{IEEE Trans.
  Inf. Theory}, vol.~61, no.~4, pp. 1795--1818, 2015.

\bibitem{gu2017information}
Y.~Gu and N.~A. Goodman, ``Information-theoretic compressive sensing kernel
  optimization and {B}ayesian {C}ram{\'e}r--{R}ao bound for time delay
  estimation,'' \emph{IEEE Trans. Signal Process}, vol.~65, no.~17, pp.
  4525--4537, 2017.

\bibitem{yeung2008information}
R.~W. Yeung, \emph{Information theory and network coding}.\hskip 1em plus 0.5em
  minus 0.4em\relax Springer Science \& Business Media, 2008.

\bibitem{burden2011numerical}
R.~L. Burden, \emph{Numerical analysis}.\hskip 1em plus 0.5em minus 0.4em\relax
  Brooks/Cole Cengage Learning, 2011.

\bibitem{barneto2019full}
C.~B. Barneto, T.~Riihonen, M.~Turunen, L.~Anttila, M.~Fleischer, K.~Stadius,
  J.~Ryyn{\"a}nen, and M.~Valkama, ``Full-duplex {OFDM} radar with {LTE} and
  5{G} {NR} waveforms: Challenges, solutions, and measurements,'' \emph{IEEE
  Trans. Microw Theory. Techn}, vol.~67, no.~10, pp. 4042--4054, 2019.

\bibitem{richards2014fundamentals}
M.~A. Richards, \emph{Fundamentals of radar signal processing}.\hskip 1em plus
  0.5em minus 0.4em\relax McGraw-Hill Education, 2014.

\end{thebibliography}

\end{document}